   \pdfoutput=1
   \documentclass[usenatbib]{mnras}
   \usepackage{graphicx}                   % Including figures
   \usepackage{times}
\newcommand{\HI}{\mathrm{H\,I}}

\newcommand{\HIa}{H\,{\sevensize{{I}}}\,\,}

\newcommand{\HeIIa}{He\,{\sevensize{{II}}}\,\,}
\newcommand{\HeIIb}{He\,{\sevensize{{II}}}}

\newcommand{\GHI}{\Gamma_{\HI}}

\newcommand{\lya}{Ly$\alpha$ }

\newcommand{\teff}{\tau_\mathrm{eff}}

\title[UVB with MFP fluctuations]{Large fluctuations in the hydrogen-ionizing background and mean free path following the epoch of reionization}
\author[F.~B. Davies, S.~R. Furlanetto]{Frederick B. Davies$^{1,2}$\thanks{davies@astro.ucla.edu}, Steven R. Furlanetto$^1$\\
$^1$Department of Physics \& Astronomy, University of California, Los Angeles, Box 951547, Los Angeles, CA 90095 \\
$^2$Max-Planck-Institut f{\"u}r Astronomie, K{\"o}nigstuhl 17, D-69117 Heidelberg, Germany}

\begin{document}

\maketitle

\begin{abstract}
Extremely large opaque troughs in the \lya forest have been interpreted as a sign of an extended reionization process below $z\sim6$. Such features are impossible to reproduce with simple models of the intergalactic ionizing background that assume a uniform mean free path of ionizing photons. We build a self-consistent model of the ionizing background that includes fluctuations in the mean free path due to the varying strength of the ionizing background and large-scale density field. The dominant effect is the suppression of the ionizing background in large-scale voids due to ``self-shielding" by an enhanced number of optically thick absorbers. Our model results in a distribution of 50 Mpc$/h$ \lya forest effective optical depths that significantly improves agreement with the observations at $z\sim5.6$. Extrapolation to $z\sim5.4$ and $z\sim5.8$ appears promising, but matching the mean background evolution requires evolution in the absorber population beyond the scope of the present model. We also demonstrate the need for extremely large volumes $(>400$~Mpc on a side) to accurately determine the incidence of rare large-scale features in the \lya forest.
\end{abstract}

\section{Introduction}

The reionization of hydrogen was an important milestone in the history of the Universe, representing the culmination of early structure formation and a dramatic phase change in the intergalactic medium (IGM). Great efforts have been made to investigate the reionization epoch, both theoretically and observationally, as it provides a powerful tool for testing theories of the formation of the first galaxies \citep{stevebook}.

One of the dominant features of neutral hydrogen is strong Lyman-series absorption. Observations of large-scale opaque regions in the \lya forest of high-redshift quasars, also known as Gunn-Peterson troughs \citep{GP1965}, have already placed intriguing constraints on the end stages of reionization (e.g. \citealt{Fan2001,Fan2006,Mesinger2010,McGreer2011,McGreer2015}). Searches for \lya damping wing absorption due to the presence of a neutral IGM \citep{ME1998} in high-redshift quasar spectra have been inconclusive (e.g. \citealt{MH2007,MF2008a,Schroeder2013,Bolton2011,Simcoe2012} but see \citealt{BB2015}), but could eventually provide ``smoking gun" evidence for primordial neutral material. \lya damping wing absorption due to a neutral IGM may also be responsible for the precipitous drop in the fraction of star-forming galaxies that show strong \lya emission lines above $z\sim6$ (e.g. \citealt{Stark2010,Ono2012,Schenker2014,Pentericci2014,Tilvi2014}), subject to considerable model-dependent uncertainties due to the complexity of \lya emission and scattering processes and inhomogeneous reionization (e.g. \citealt{Santos2004,Furlanetto2004,McQuinn2007,MF2008b,Dijkstra2011,BH2013,Choudhury2014,Mesinger2015}).

The ionization state of gas in the IGM is regulated by the ``ionizing background", the extragalactic ionizing radiation field. The ionizing background is comprised of ionizing photons emitted by young stars and quasars, filtered by neutral gas structures that set an average attenuation length or mean free path $\lambda$ (e.g. \citealt{HM1996,HM2012,FG2009}). Interpretation of \lya forest observations typically assumes a uniform hydrogen ionization rate $\GHI$ due to the large mean free path ($\lambda>100$ Mpc) of ionizing photons at $z < 5$ that smooths the radiation field on large scales. More sophisticated models include the effect of discrete sources of ionizing photons and attenuate ionizing radiation with a constant mean free path (e.g. \citealt{MW2004,McDonald2005,WL2006,MF2009}). Recently, \citet{Becker2015} (henceforth B15) discovered an enormous $\sim100$ Mpc$/h$ opaque trough in the \lya forest spectrum of ULAS J0148+0600 covering $z\sim5.5-5.9$ which is at odds with their uniform mean free path ionizing background simulations. They suggest that this discrepancy could be due to fluctuations in the mean free path of ionizing photons, possibly as a result of an extended reionization process.

However, the mean free path of ionizing photons is likely to vary substantially at $z\sim5.6$ even if ionization equilibrium is assumed. Post-processing of hydrodynamical simulations of the IGM suggests that the mean free path should vary with the strength of the ionizing background as $\lambda\propto\GHI^{\xi}$ with $\xi\sim0.6$--$0.8$ (\citealt{McQuinn2011}; henceforth M11), and uniform mean free path ionizing background simulations at $z\sim5.6$ lead to large-scale (tens of Mpc) regions with enhanced and diminished background \citep{MF2009}. Regions with a weak ionizing background should thus have a short mean free path, while regions with a strong ionizing background should be more transparent to ionizing photons. This effect causes a feedback loop that should lead to enhanced contrast of the ionizing background on (roughly speaking) scales larger than the average mean free path of ionizing photons. Fluctuations in the mean free path have recently been studied with linear theory at $z\sim2.5$ \citep{Pontzen2014,Gontcho2014} in terms of subtle modulation of the baryon acoustic oscillations signal, but have been ignored at higher redshift where they are potentially very important to the large-scale structure of the radiation field.

In this work, we construct a semi-numerical model to compute the ionizing background in a large cosmological volume at $z\sim5.6$ including a varying mean free path of ionizing photons. The local mean free path is iteratively determined by the local overdensity and strength of the ionizing background, the latter of which is computed by filtering ionizing radiation through the varying opacity field. The resulting radiation field is ``self-consistent" in the sense that the opacity of the IGM and ionizing background are set by each other -- the opacity depends on the strength of the ionizing background, and the ionizing background itself is filtered by that opacity. We then investigate the statistics of large-scale features in \lya forest transmission that our new ionizing background model implies and compare to the observations compiled by B15.

The structure of the paper is as follows. In Section~\ref{sec:sourcesink}, we describe our models for sources and absorbers of ionizing photons. In Section~\ref{sec:bkgdmodel}, we compute the self-consistent fluctuations in the ionizing background and mean free path. In Section~\ref{sec:hilyaforest}, we apply our ionizing background model to the statistics of large-scale \lya forest absorption. In Section~\ref{sec:zevol}, we test the redshift evolution of our model across $z\sim5.8$--5.4. In Section~\ref{sec:hidiscuss}, we discuss the implications of our model for the IGM and sources of ionizing photons at high-redshift. Finally, in Section~\ref{sec:hiconclude} we conclude with a summary and speculate on future observational tests of our model.

In this work we assume a standard $\Lambda$CDM model with $\Omega_\mathrm{m}=0.3$, $\Omega_\mathrm{\Lambda}=0.7$, $h = 0.7$, and $\sigma_8=0.82$, consistent with constraints from the cosmic microwave background (\citealt{Hinshaw2013}, and close to the latest bounds by \citealt{Planck2015}). Distance units are comoving unless otherwise noted.

\section{Sources and Sinks of Ionizing Photons}\label{sec:sourcesink}

Our three-dimensional model for the ionizing background requires two ingredients: the distribution of ionizing sources and a prescription for spatially variable IGM opacity to ionizing photons (i.e. a varying mean free path). 

\subsection{Semi-numerical Density and Halo Fields with {\small{DEXM}}}\label{sec:dexm}

\begin{figure*}
\begin{center}
\resizebox{18cm}{!}{\includegraphics{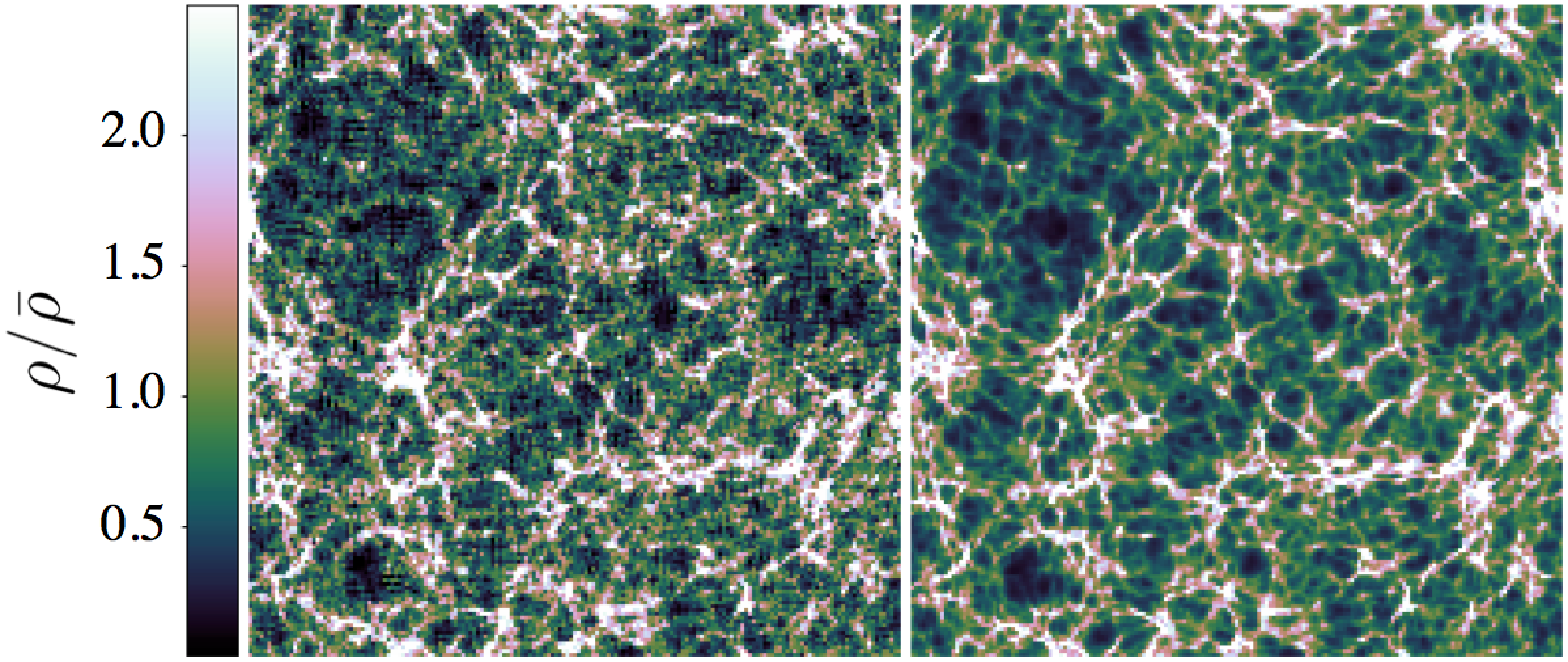}}\\
\end{center}
\caption{Comparison of the standard nearest-gridpoint Zel'dovich approximation density field from {\scriptsize{DEXM}} (left) with our interpolated Zel'dovich velocity approach (right). The slice shown is 0.57 Mpc thick (1 pixel) and 100 Mpc on a side.}
\label{fig:denscompare}
\end{figure*}

To construct the distribution of ionizing sources in our model, we use the semi-numerical cosmological simulation code {\small DEXM} (\citealt{MF2007}; henceforth MF07). In the following, we summarize the method employed in {\small{DEXM}}, deferring a discussion of the detailed implementation to MF07. We begin with a 2100$^3$ grid of Gaussian cosmological initial conditions (ICs) in a volume 400 Mpc on a side. While this volume is considerably larger than the majority of cosmological simulations, we will investigate whether this is a representative volume in Section \ref{sec:cv}. The linear velocity field is then computed on a coarser grid of 700$^3$ cells using the Zel'dovich approximation (\citealt{Zeldovich1970,Efstathiou1985}). In the public release of {\small{DEXM}}, the quasi-linear density field is then computed by displacing ``particles," corresponding to the grid of initial conditions, following the coarsely-computed velocity field (with no interpolation) and assigning their mass to the nearest grid cell to compute the density field. The right panel of Figure~\ref{fig:denscompare} shows that while the standard {\small DEXM} density field is reasonably smooth in high-density regions with a large number of particles per cell, low-density regions suffer from considerable shot noise. In particle-based cosmological simulations, a nearest-neighbor smoothing approach is typically performed on the particle distribution to estimate the continuous density field (e.g. \citealt{Monaghan1992,Springel2005}). Nearest-neighbor schemes naturally degrade resolution in low-density environments in favor of high-density regions, ideal for computationally intensive hydrodynamical simulations of galaxy formation. However, these void environments are crucial to modeling transmission through the \lya forest at $z>5$ \citep{BB2009}, so we adopt a novel approach that does not degrade our simulation resolution through explicit spatial smoothing.

The linear velocity field produced by the Zel'dovich approximation is typically smooth on small scales, its structure dominated by large-scale bulk flows of material into sheets and filaments. To compute the quasi-linear density field, we displace a super-resolved grid of 21000$^3$ IC particles\footnote{The required number of sub-particles to produce a qualitatively smooth (and converged) final density field depends on the relative number of IC vs. velocity grid cells, and on the spatial resolution of the output density field.} by interpolating the Zel'dovich velocity field via trilinear interpolation and then binning these sub-particles onto a 700$^3$ grid.\footnote{Matching the resolution of the Zel'dovich velocity field is not strictly required -- one could in principle bin the sub-particles onto a coarser or finer grid. In the latter case, the resulting density field appears to have subtle box-like artifacts, presumably due to the linear interpolation of the velocity field.} The material in each original IC grid cell is not only displaced but also stretched and sheared due to gradients in the velocity field. Because the velocity field is smooth on small scales, this procedure results in a smooth distribution of matter on a uniform grid without any additional spatial filtering -- even in the lowest density environments. The left panel of Figure~\ref{fig:denscompare} shows the resulting density field computed from the same ICs as the right panel, demonstrating the lack of artifacts -- especially in void environments -- compared to the standard method. 

\begin{figure}
\begin{center}
\resizebox{8cm}{!}{\includegraphics{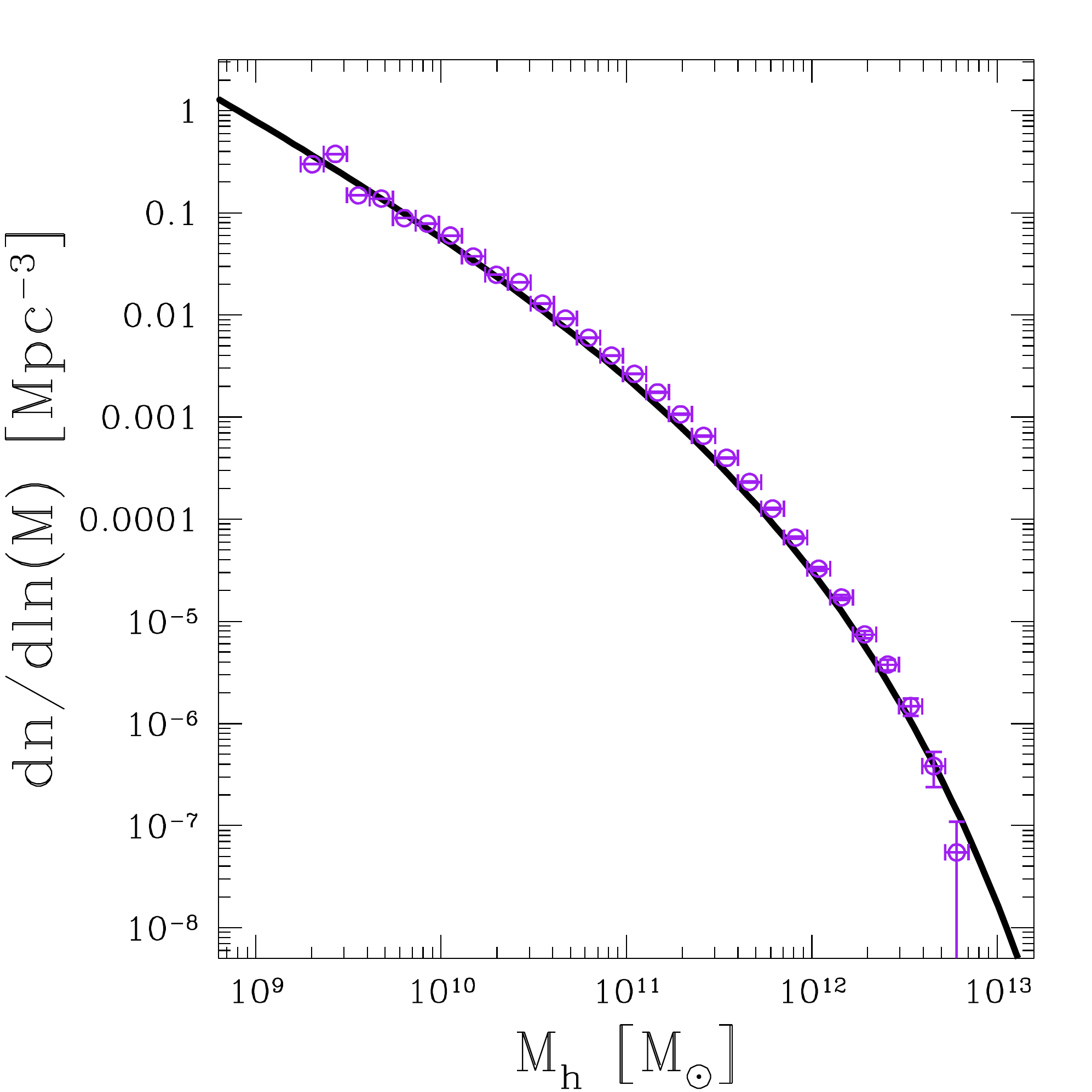}}\\
\end{center}
\caption{\citet{ST1999} halo mass function at $z=5.6$ (solid curve) compared to the mass function of halos in our {\scriptsize{DEXM}} simulation (open points).}
\label{fig:hmf}
\end{figure}

We use the halo finding algorithm of {\small{DEXM}} to populate the simulation volume with a realistically-clustered distribution of dark matter halos. In brief, halos are located by filtering the linear density field on successively smaller scales and searching for (non-overlapping) regions where the filtered density $\delta(\vec{x},M)$ is greater than the linear collapse threshold $\delta_c$. We locate halos with masses as low as $M_h \sim 2\times10^9 M_\odot$ ($\sim30$ IC cells) and show the resulting mass function at $z=5.6$, which (by construction) agrees very well with the \citet{ST1999} mass function in Figure~\ref{fig:hmf}. We then displace the halos using the Zel'dovich velocity field described above, enhancing their bias on large-scales and resulting in clustering statistics that are consistent with N-body simulations (MF07). As discussed by MF07, we do not expect one-to-one agreement between halo locations found with this procedure and those formed in an N-body simulation with the same initial conditions, but the agreement with large-scale clustering statistics is adequate for our purposes.

To compute the ionizing photon output of each halo, we abundance match (e.g. \citealt{VO2004}) to the observed UV luminosity functions from \citet{Bouwens2015b}, interpolating between their $z\sim5$ and $z\sim6$ best-fit Schecter functions and assuming a constant ratio of $f_\mathrm{ion}/A_{912}$ between the non-ionizing and ionizing continuum luminosity as in B15, where $A_{912}=6.0$ is the expected ratio for a young stellar population and $f_\mathrm{ion}$ is a free parameter that represents a combination of the escape fraction of ionizing photons $f_\mathrm{esc}$ and uncertainties in the value of $A_{912}$. For comparison to the B15 models we further assume that the spectrum of ionizing photons emitted by each galaxy is a power law $L_\nu \propto \nu^{-\alpha}$ with $\alpha=2.0$.\footnote{See Section 5.1 of \citet{BB2013} for a detailed assessment of this choice.}

\subsection{Varying Mean Free Path}\label{sec:fluctmfp}

The mean free path of ionizing photons in the IGM has been measured out to $z\sim5$ through stacking analyses of quasar ionizing continua (Worseck et al. 2014) and by counting absorption lines in the \lya forest (e.g. \citealt{SC2010,Rudie2013}). Typically this mean free path is assumed to be uniform in space. However, the mean free path itself should be sensitive to the strength of the ionizing background due to the regulation of \HIa absorber sizes (M11; see also \citealt{Munoz2015}) and overdense regions likely contain more dense self-shielded gas.

M11 developed an analytic framework to describe the self-consistent relationship between the mean free path $\lambda$ and the ionization rate $\Gamma$ based on the IGM model of \citet{ME2000} (henceforth MHR). The MHR model describes the ionization state of the universe by assuming that all gas with density above a critical value $\Delta_i$ is neutral, self-shielded from the ionizing background, while all lower density material is completely ionized.\footnote{In detail, a substantial amount of IGM ionizing opacity should come from optically thin absorbers. However, the \emph{scaling} of $\lambda$ with $\Gamma$ should not be sensitive to this assumption, as confirmed in numerical simulations by M11.} This means that global parameters that depend on the amount of neutral or ionized gas can be completely specified through the gas density probability distribution function (PDF) $P(\Delta)$. Additionally, the MHR model assumes that the mean free path is given by $\lambda \propto F_V(\Delta_i)^{-2/3}$, where $F_V(\Delta_i)$ is the volume filling factor of gas above the density $\Delta_i$. If the gas density PDF can be approximated by a power law $P(\Delta)\propto\Delta^{-\gamma}$ in the density regime that dominates the IGM opacity, then $\lambda \propto \Delta_i^{2(\gamma-1)/3}$. Under the assumption of ionization equilibrium, M11 showed that $\Gamma\propto\Delta_i^{(7-\gamma)/3}$. In other words, $\lambda\propto\Gamma^{2(\gamma-1)/(7-\gamma)}$. Fundamentally, if the ionization rate is larger, higher density gas will be ionized, reducing the total volume of neutral gas and thus increasing the mean free path (see also \citealt{Munoz2015}). M11 then confirmed this scaling by performing radiative transfer through hydrodynamic cosmological simulations.

In our fiducial simulations we allow the mean free path to vary as $\lambda\propto\Gamma^{2/3}$, corresponding to a gas density PDF $P(\Delta)\propto\Delta^{-2.5}$. This PDF is equivalent to assuming that all gas lies in spherically symmetric clouds with an isothermal density profile, i.e. $\rho\propto{r}^{-2}$. This is likely a conservative assumption; M11 found that the true PDF is likely to be steeper at high redshift, leading to a more sensitive dependence with $\GHI$. In Section~\ref{sec:hidiscuss} we will discuss the effect of modifying this choice.

The variation of the mean free path with large-scale density depends on the relative number density of self-shielded absorbers as a function of that large-scale density. In general, this variation should moderate the amplitude of mean free path fluctuations because (if absorbers are biased relative to dark matter) it acts in the opposite direction from the $\Gamma$ dependence described above (in that voids will become more transparent). At $z\ga5.5$, the overdensity threshold for self-shielding is less than the virial overdensity, and so the population of absorbing clouds is likely dominated by the distant outskirts of very low-mass halos and filamentary structures in the IGM \citep{Munoz2015}. These overdense structures are unresolved in our simulations and so we only attempt to describe the variation in their abundance on the same coarse scale as the ionizing background (see below). The absorbing clouds are likely only weakly biased with respect to the density field, so we assume a fiducial absorption bias of unity for simplicity (i.e. $\lambda\propto\Delta^{-1}$). However, in the densest environments in our simulations which have the strongest ionizing background, the absorber population may instead be dominated by gas within the virial radius of low-mass halos \citet{Munoz2015}, leading us to overestimate the mean free path in these regions. We leave a full exploration of the relationship between large-scale overdensity and the mean free path for future work.

For comparison with B15, we assume that the column density distribution of absorbers that are most relevant to \HIa opacity ($N_\HI\sim N_\mathrm{LL}\sim10^{17-18}$ cm$^{-2}$) behaves as a power law $f(N_\HI)\propto N^{-\beta}$ with $\beta=1.3$, consistent with the $z\sim2$--$5$ model in \citet{BB2013}. This implies a frequency dependence of the mean free path of $\lambda_\nu \propto \nu^{3(\beta-1)} \propto \nu^{0.9}$ (e.g. \citealt{HM1996}). In practice, our results are insensitive to this choice because of the already steep frequency dependence of the \HIa cross section ($\propto \nu^{-3}$) and galaxy spectrum ($\propto \nu^{-2}$) which also contribute to the ionization rate calculation in the following section.

The final expression that we use for the mean free path is then:
\begin{equation}
\lambda(\Gamma,\Delta,\nu) = \lambda_0 (\GHI/\Gamma_0)^{2/3} \Delta^{-1} (\nu/\nu_\HI)^{0.9},
\end{equation}
where $\lambda_0$ and $\Gamma_0$ are tuned, via re-normalizing the ionizing emissivity with $f_\mathrm{ion}$, such that an ionizing background model with a spatially constant $\lambda=\lambda_0$ produces $\langle \GHI \rangle = \Gamma_0$.

Extrapolation of the \citet{Worseck2014} mean free path evolution fit (derived from measurements over $z\sim2.5$--$5.2$) to $z\sim5.6$ suggests $\lambda \sim 54$ Mpc as an estimate of the average mean free path. However, the ionization rate in the IGM is constant from $2 \la z \la 5$ \citep{BB2013} before dropping by roughly an order of magnitude from $z\sim5$ to $z\sim6$ \citep{BH2007a,WB2011,Calverley2011}. This suggests that extrapolation of this fit to higher redshifts is likely to \emph{overestimate} the mean free path by as much as a factor of a few. In this work we are interested in exploring a range of fluctuating background scenarios, so we investigate models with $\lambda_0=$ 15, 22, and 34 Mpc.

\section{Numerical model of the ionizing background}\label{sec:bkgdmodel}

In this section, we describe the method used to compute the ionizing background given the ingredients listed in the previous section.

Our fiducial ionizing background models are computed on a coarse 80$^3$ grid, corresponding to cells of 5 Mpc on a side. This coarse resolution is sufficient for our purposes because it resolves the typical mean free path in our simulations (which sets the smoothing scale) and we have found little difference in the large-scale radiation field structures over a range of grid sizes from 48$^3$ to 100$^3$. We also bin the density field onto the same coarse grid for computational efficiency -- as described in the previous section, we do not resolve neutral gas in our simulations, but instead assume that the number density of absorbers scales proportional to the large-scale overdensity. The halo-finding procedure in Section~\ref{sec:dexm} results in $\sim2.4\times10^7$ halos. While a uniform mean free path calculation of the radiation field is computationally trivial for any number of sources (i.e. $J_\nu\propto\sum L_\nu\mathrm{e}^{-R/\lambda}/R^2$, or computed via FFT), self-consistently including the mean free path variations from Section~\ref{sec:fluctmfp} requires line-of-sight integrations between each source and grid cell. For computational efficiency, we bin the ionizing sources from Section~\ref{sec:dexm} -- originally found on a 2100$^3$ grid -- onto the ionizing background grid, then compute the local and non-local contributions to the ionizing background separately.\footnote{This reduces computation time in two ways: the number of sources drops from $\sim2.4\times10^7$ to $\sim5\times10^5$, and the opacity between each pair of cells only has to be computed once. Thus, the total computation time is decreased by a factor of $\sim100$. The gain in efficiency depends on the resolution of the 3D grid.}

The non-local contribution to the specific intensity $J_\nu$ at cell $i$ is summed over every other cell $j$ located at $\vec{r}_j$,
\begin{equation}\label{eqn:jnu}
J_{\nu,i} = \sum_{j{\neq}i} \frac{L_{\nu,j}}{(4\pi|\vec{r}_i-\vec{r}_j|)^2} \mathrm{e}^{-\tau_{\nu}(\vec{r}_i,\vec{r}_j)},
\end{equation}
where $L_{\nu,j}$ is the specific luminosity of cell $j$ and $\tau_{\nu}(\vec{r}_i,\vec{r}_j)$ is the integrated optical depth between cells $i$ and $j$,
\begin{equation}
\tau_{\nu}(\vec{r}_i,\vec{r}_j) = \int_{\vec{r}_i}^{\vec{r}_k} [\lambda_\nu(\GHI(\vec{r}),\Delta(\vec{r}))]^{-1} dr,
\end{equation}
where $\lambda_\nu$ is the mean free path of frequency $\nu$ photons corresponding to $\GHI$ and $\Delta$ along the line of sight. By integrating along the line of sight, the varying \HIa opacity of the IGM due to fluctuations in the strength of the radiation field and the large-scale density field is explicitly taken into account as described in the previous section.\footnote{In this simple model we neglect cosmological effects, most notably the redshifting of ionizing photons as they travel from the source cell. Fortunately, the assumed mean free path of ionizing photons is short enough that this is unlikely to have a significant effect on our results.}

The local contribution is computed by assuming that sources are spread evenly throughout each cell and can be described with a uniform emissivity $\epsilon_\nu$. The radiation field at the center of a uniform emissivity sphere of radius $l/2$ and mean free path $\lambda$ is simply the integral over spherical shells with luminosity $dL~=~\epsilon \times 4\pi r^2 dr$ or
\begin{equation}
J_\nu^{\mathrm{local},\mathrm{sphere}} = \int_0^{l/2} \frac{\epsilon_\nu (4\pi r^2)} {(4\pi r)^2} \mathrm{e}^{-r/\lambda_\nu} dr = \frac{\epsilon_\nu\lambda_\nu}{4\pi} (1-\mathrm{e}^{-(l/2)/\lambda_\nu}),
\end{equation}
which reverts to the ``absorption limited" approximation for the ionizing background intensity $J \approx \epsilon\lambda/4\pi$ \citep{MW2003} when integrated out to infinity. There is no simple exact expression for the radiation field at the center of a cubical volume with uniform emissivity, but we find numerically that the approximation
\begin{equation}
J_\nu^{\mathrm{local},\mathrm{cube}} \approx \frac{\epsilon_\nu\lambda_\nu}{4\pi} (1-\mathrm{e}^{-\zeta{l}/\lambda_\nu})
\end{equation}
with $\zeta \approx 0.72$ results in an average ionization rate that is independent of spatial resolution.

Finally, the hydrogen ionization rate $\Gamma_\HI$ in each cell is computed by integrating over frequency,
\begin{equation}
\Gamma_{\HI,i} = 4\pi\int_{\nu_\HI}^{4\nu_\HI} \frac{J_{\nu,i} + J_{\nu,i}^{\mathrm{local}}}{h\nu} \sigma_\HI(\nu) d\nu,
\end{equation}
where $\sigma_\HI(\nu)$ is the hydrogen photoionization cross-section from \citet{Verner96}. This additional integration step considerably increases the computation time of our simulations relative to uniform mean free path models. By testing with lower-resolution 48$^3$ models, we find that the ionizing background fluctuations in the frequency-integrated model can be well-reproduced by a single-frequency model with $\nu \approx 1.32 \nu_\HI$, and use this approximation to speed up computation of our highest resolution models.\footnote{The 80$^3$ single-frequency models presented here require approximately 7 hours of runtime per iteration on a 12-core Intel Xeon (Nehalem generation) computer.}

Because the local mean free path is a function of $\GHI$, the above procedure must be iterated $\sim10$ times until the average change of $\GHI$ per cell is less than $0.3\%$. At this point the ionization rate in the vast majority of the volume has converged, except for a very small fraction of cells at the center of opaque void regions that are strongly shielded from all ionizing sources.

In the rest of the paper, we will describe a series of models with a varying average mean free path. Models that employ a uniform mean free path (i.e. $\lambda_\nu = $ constant in equations 2 and 3 above) will be referred to as ``uniform-$\lambda$" while those that include our full prescription for mean free path variations will be referred to as ``fluctuating-$\lambda$". The uniform-$\lambda$ models have been normalized to reproduce a single average $\Gamma$ by tuning the normalization of the ionizing emissivity through $f_\mathrm{ion}$ (analogous to changing the assumed escape fraction of ionizing photons), and the fluctuating-$\lambda$ models were computed using the uniform-$\lambda$ model as the first iteration. Throughout most of the paper, we will utilize a single realization of the density field (with a volume of $[400 \ {\rm Mpc}]^3$) in order to illuminate the dependence of the results on the absorber properties.  We will examine the role of cosmic variance in Section~\ref{sec:cv} separately, but we note that our fiducial realization appears to be a \emph{conservative} estimate of the impact of a fluctuating mean free path.

\begin{figure}
\begin{center}
\resizebox{8cm}{!}{\includegraphics{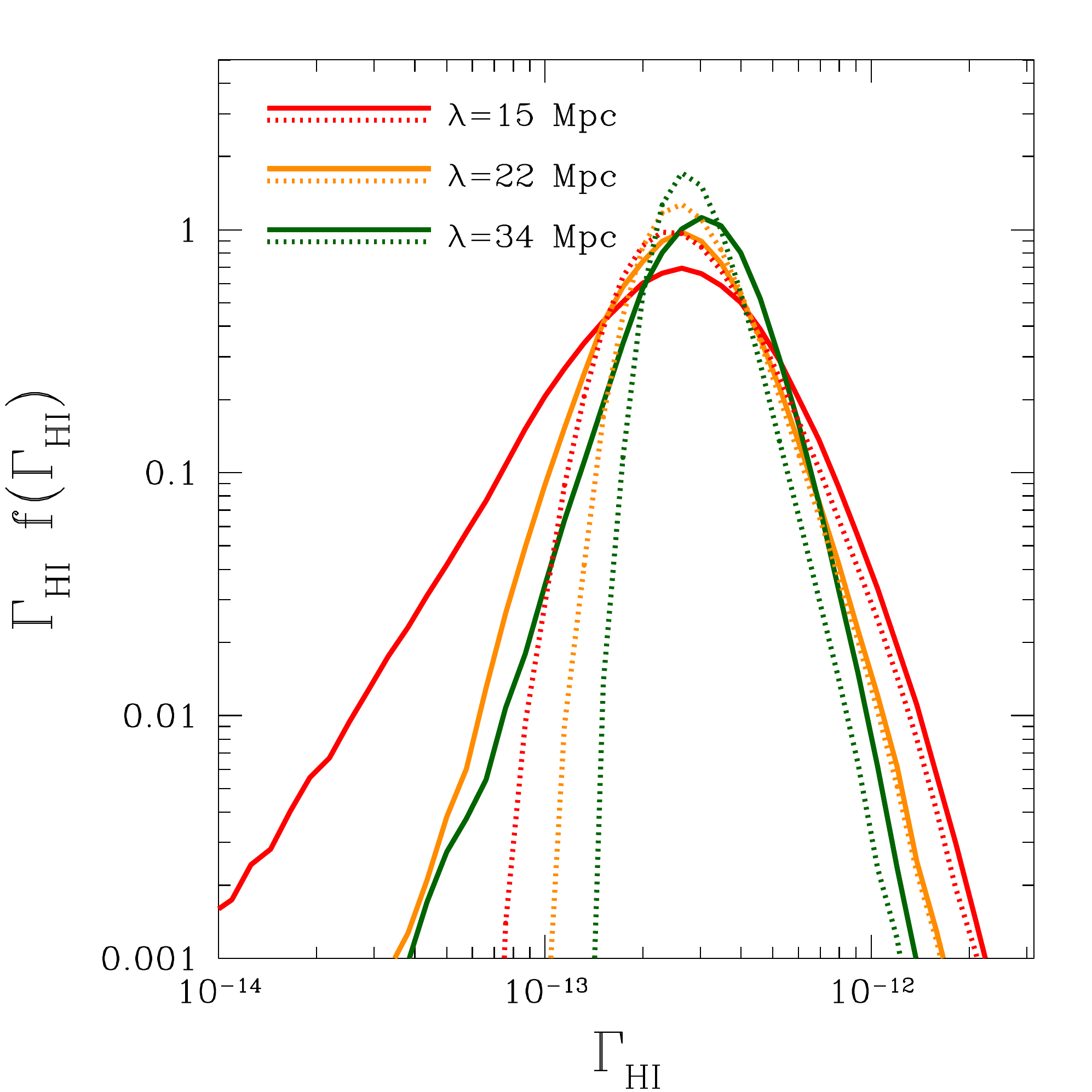}}\\
\end{center}
\caption{The solid curves show $\GHI f(\GHI)$ from the fiducial ionizing background simulation at $z=5.6$ for $\lambda=15,22,34$ Mpc in red, blue, and orange, respectively. The dotted curves show the distributions from the corresponding uniform MFP models (i.e. the first iteration of the ionizing background calculation). The differences between the two sets therefore demonstrate the effect of mean free path fluctuations.}
\label{fig:fgammahi}
\end{figure}

\begin{figure*}
\begin{center}
\resizebox{16cm}{!}{\includegraphics{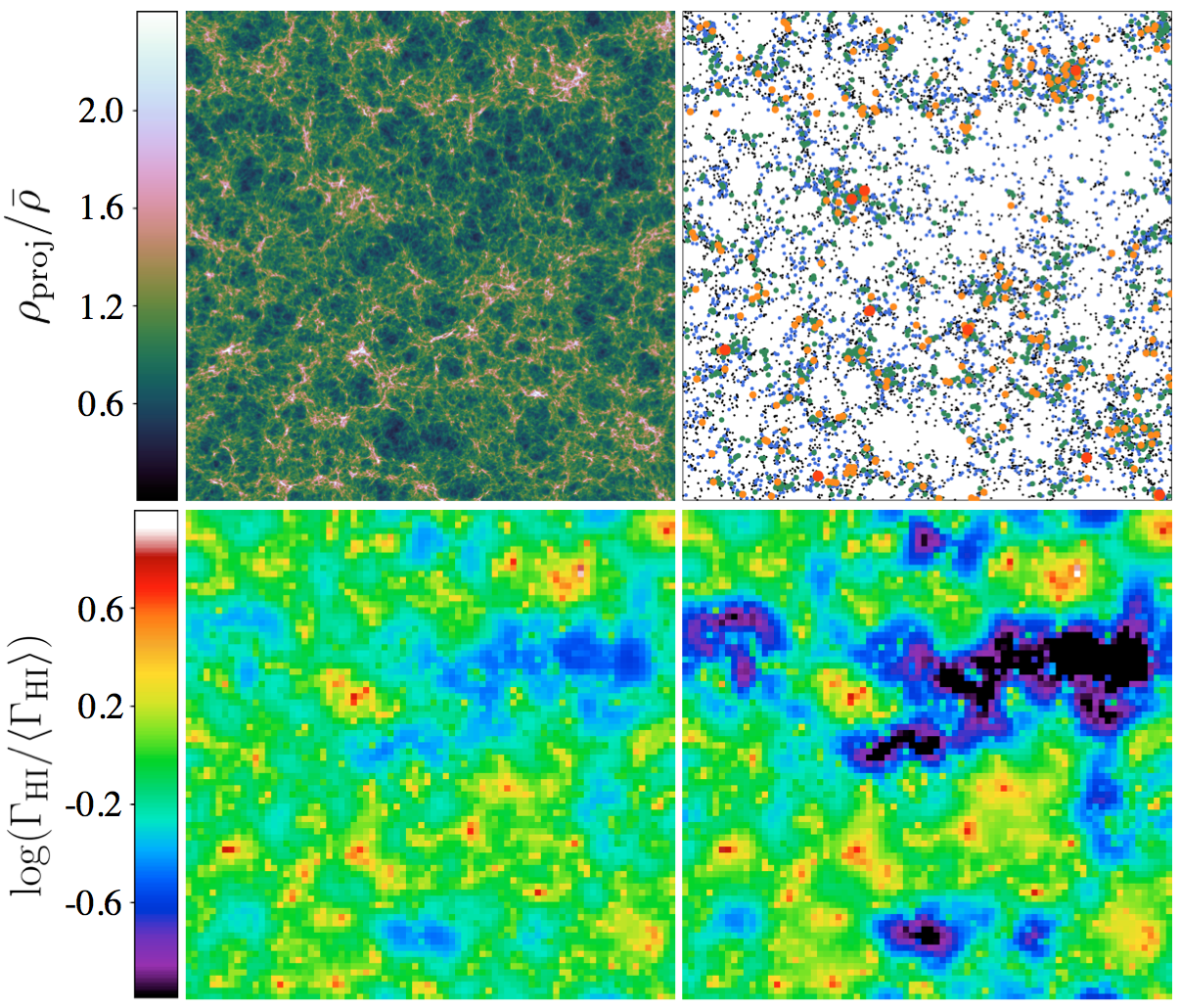}}\\
\end{center}
\caption{Top left: 20 Mpc-thick slice of the density field in our fiducial cosmological volume, 400 Mpc on a side. Top right: Halos found using the {\scriptsize DEXM} halo-finding procedure inside the same slice. The size and color of each point represents the corresponding UV magnitude of the halo when matched to the \citet{Bouwens2015b} luminosity function, from $M_\mathrm{UV}\sim-18$ (black, smallest) through $M_\mathrm{UV}\sim-22$ (red, largest) in steps of $\Delta M=1$. The halos shown represent the brightest $\sim1\%$ of all galaxies in this slice, with halo masses ranging from $\sim10^{10.8}$--$10^{12.3} M_\odot$. Bottom left: Fluctuations in the uniform-$\lambda=15$ Mpc ionizing background model in a 5 Mpc-thick slice corresponding to the center of the slice shown in the density and halo fields. Variations of a factor of $\sim2$--3 are common on large scales, as found by previous authors. Bottom right: Fluctuations in the fluctuating-$\lambda=15$ Mpc ionizing background model in the same 5 Mpc-thick slice. The addition of mean free path fluctuations greatly increases the fluctuations in $\GHI$, especially to very low values in large-scale underdensities.}
\label{fig:denshalouvb}
\end{figure*}

\begin{figure}
\begin{center}
\resizebox{8cm}{!}{\includegraphics{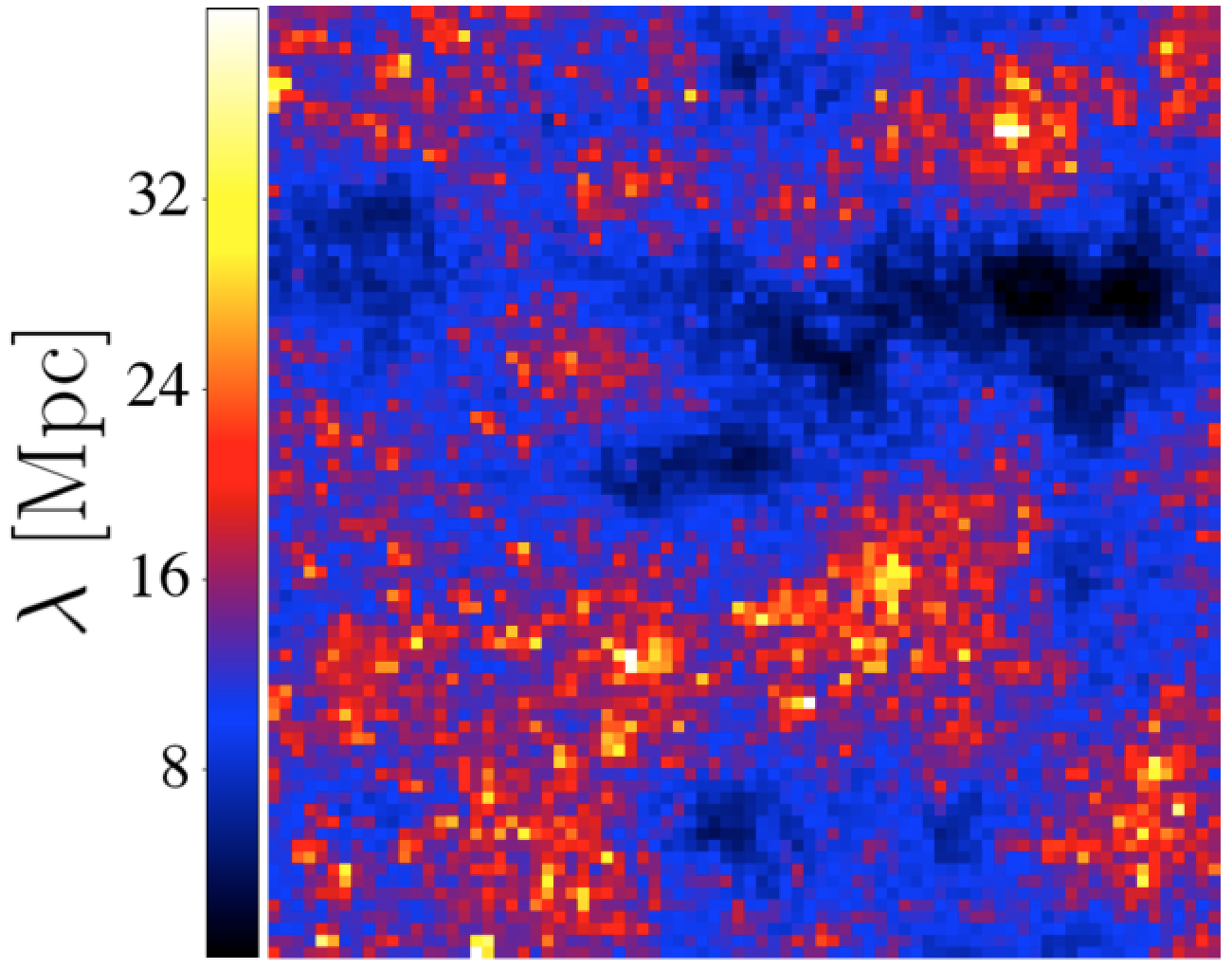}}\\
\end{center}
\caption{Mean free path of hydrogen-ionizing photons in the ionizing background model slice shown in the bottom right panel of Figure~\ref{fig:denshalouvb}, where $\lambda\propto\GHI^{2/3}\Delta^{-1}$ and $\lambda(\langle\GHI\rangle)=15$ Mpc. The strong fluctuations in $\GHI$ dominate the large-scale features, while small-scale ``noise" is due to the relatively small variations in $\Delta$.}
\label{fig:mfp}
\end{figure}

\subsection{Ionizing Background Results}

In Figure~\ref{fig:fgammahi}, we show the distribution of $\Gamma_\HI$ for our fiducial fluctuating-$\lambda$ models assuming $\lambda=15$, 22, and 34 Mpc. The solid curves show the self-consistent fluctuating-$\lambda$ models, while the dotted curves show the corresponding uniform-$\lambda$ models. Mean free path fluctuations substantially broaden the distribution of $\GHI$ towards lower values, and the strength of the effect strongly depends on the average $\lambda$. This tail to low $\GHI$ comes about in large-scale underdensities where the weak $\GHI$ -- and thus, the shorter $\lambda$ -- acts to ``self-shield" the region from distant sources of ionizing photons which would otherwise lead to a more uniform, shallower trough in $\GHI$. The high-$\GHI$ end of the distribution is only modestly adjusted for two reasons. First, the density and ionization adjustments to the mean free path partially cancel out, leading to only a modest boost in the mean free path. Second, close to galaxies the structure of the radiation field is dominated by the $1/r^2$ term in $J_\nu$ (equation \ref{eqn:jnu}), so increasing the mean free path has a lesser effect than one might expect (i.e. $J_\nu\propto\lambda$, see \citealt{MW2003}).

Figure~\ref{fig:denshalouvb} shows density, halo, and ionizing background fields centered on the one slice of our fiducial model. The bottom panels demonstrate the spatial coherence of the background fluctuations with 5 Mpc (1 pixel) thick slices of the $\lambda=15$ Mpc model in the uniform-$\lambda$ (left) and fluctuating-$\lambda$ (right) cases. The addition of a fluctuating mean free path greatly increases the contrast of the radiation field on $\ga\lambda$ scales. The upper panels of Figure~\ref{fig:denshalouvb} show projected density and halo fields $\pm10$ Mpc from the center of the ionizing background slice. Comparison to the bottom panels shows that the spatial fluctuations in our ionizing background model are due to the inhomogeneous distribution of ionizing sources related to the formation of large-scale structure. Figure~\ref{fig:mfp} shows the fluctuating mean free path in a slice corresponding to the lower right panel of Figure~\ref{fig:denshalouvb}. In this case, $\lambda$ represents the ``effective" mean free path inside each UVB pixel, i.e. the optical depth to hydrogen ionizing photons across the pixel is $\tau_\HI = dR/\lambda$ where $dR$ is the path length through the pixel. Large-scale features in the radiation field dominate the structure of mean free path map, with small-scale ``noise" reflecting variations in the density field on the 5 Mpc pixel scale. 

\section{Implications for Large-scale \lya Forest Transmission}\label{sec:hilyaforest}

Our simulations do not resolve the density and velocity fields at the Jeans scale of the gas at these redshifts, so we cannot produce realistic \lya forest spectra. We can still estimate the effect the ionizing background fluctuations on the large-scale opacity of the \lya forest by employing the fluctuating Gunn-Peterson approximation (FGPA; e.g. \citealt{Weinberg1997}) on sightlines through our evolved density field,
\begin{eqnarray}\label{eqn:taugp}
\tau_\mathrm{GP} \approx 35.6\kappa\,\left( \frac{T_0}{7500\,\mathrm{K}} \right)^{-0.724} \left( \frac{\GHI}{3\times10^{-13}\,\mathrm{s}^{-1}} \right)^{-1} \nonumber \\
\times\,\Delta^{2-0.724(\gamma-1)} \left( \frac{1+z}{6.6} \right)^{4.5},
\end{eqnarray}
where $T_0$ is the temperature of the IGM at the mean density, $\gamma$ is the slope of the temperature-density relation, and $\kappa$ is a normalization factor (e.g. \citealt{DF2009}) that we adjust to match the observed distribution of \lya forest optical depths, similar to B15\footnote{In detail, B15 instead varied $\GHI$ in their simulations to match observations, but in practice the behavior is identical.}. We assume $T_0=7500$ K and $\gamma=1.5$, consistent with standard models for the thermal history of the IGM\footnote{Such models typically assume that reionization is complete by $z\ga9$. If instead reionization did not end until $z\sim6$, the thermal state of the IGM could be very different. The principal effect would be to increase the mean temperature and hence shift the required normalization factor $\kappa$ closer to unity.} (e.g. \citealt{Puchwein2015}). We choose a characteristic scale of 50 Mpc$/h$ for comparison to B15 and compute the effective optical depth $\teff = -\mathrm{ln}(\sum \mathrm{e}^{\tau_\mathrm{GP}(r_i)}/N)$ in steps of $\sim0.06$~Mpc along each sightline. Our fiducial simulations have $\langle\GHI\rangle\sim3\times10^{-13}$ s$^{-1}$ and require $\kappa\sim0.14$ to match the $P(\teff)$ observations by B15. The need for such a small value of $\kappa$ is primarily due to the inability of our simulations to resolve the low-density gas in small-scale cosmic voids which dominates the transmission in the relatively opaque $z>5$ \lya forest \citep{BB2009}. For the rest of this work, we assume that the \emph{dependence} of $\teff$ on $\GHI$ is captured by our relatively low resolution simulation and note that we cannot place any constraints on the true mean value of $\GHI$.

\begin{figure*}
\begin{center}
\resizebox{18cm}{!}{\includegraphics{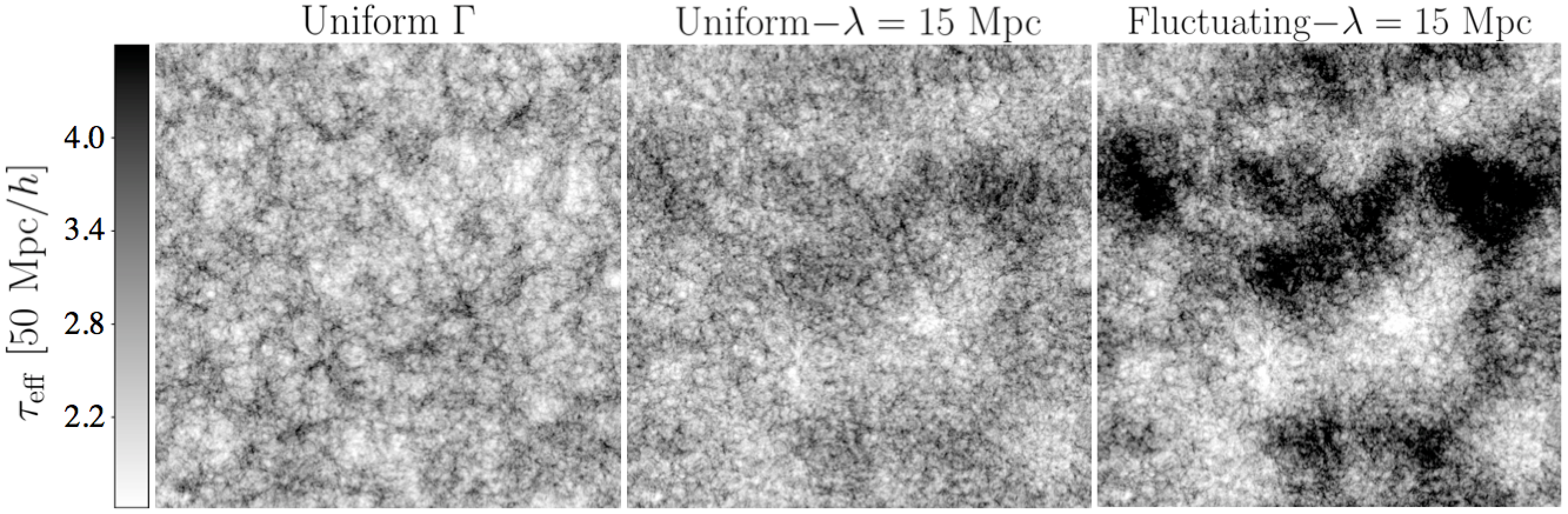}}\\
\end{center}
\caption{Maps of the 50 Mpc$/h$-projected $\teff$ in the \lya forest at $z=5.6$ centered on the same slice of the density field shown in Figure~\ref{fig:denshalouvb}. The left panel shows the $\teff$ map for a uniform ionizing background, where the opacity is correlated with the density field (see the upper left panel in Figure~\ref{fig:denshalouvb}). The middle panel demonstrates the effect of including a fluctuating ionizing background with a uniform $\lambda$: the relationship between projected density and $\teff$ reverses on large-scales due to the correlation of the density field with the radiation field. The right panel includes the full fluctuating-$\lambda$ ionizing background model, greatly increasing the correlation and leading to very high $\teff$ in the center of large-scale voids.}
\label{fig:lyamaps}
\end{figure*}

In Figure~\ref{fig:lyamaps} we show ``maps" of \lya forest opacity along 50 Mpc$/h$ sightlines perpendicular to the page, centered on the same slice shown in Figure~\ref{fig:denshalouvb}, for three different models: uniform $\Gamma$, uniform-$\lambda$=15 Mpc, and fluctuating-$\lambda$=15 Mpc from left to right. It is clear that coherent large-scale structure in the radiation field at this redshift plays an important role in the large-scale opacity of the \lya forest, and that the addition of mean free path fluctuations greatly enhances this effect -- especially on scales larger than the average mean free path itself.

\begin{figure}
\begin{center}
\resizebox{8cm}{!}{\includegraphics{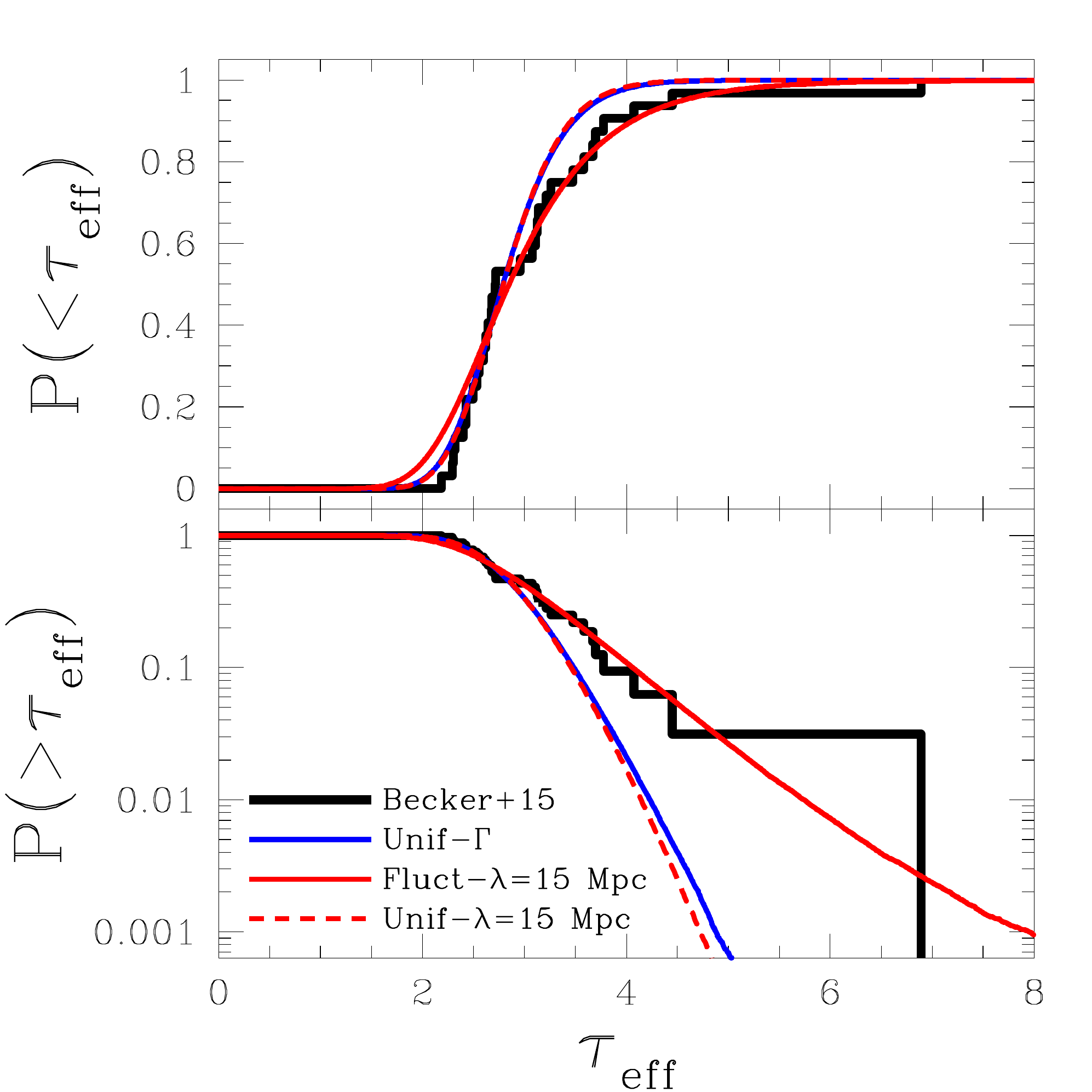}}\\
\end{center}
\caption{Top: Cumulative optical depth distribution $P(<\teff)$ computed for the uniform background (blue curve) and $\lambda=15$ Mpc fluctuating background models with uniform-$\lambda$ (dashed red) and fluctuating-$\lambda$ (solid red). The observed distribution from \citet{Becker2015}, including lower limits, is shown as the black curve. Bottom: The same curves as the top panel but recast as $P(>\teff)=1-P(<\teff)$ and shown on a logarithmic scale to emphasize the high-$\teff$ behavior.}
\label{fig:taudist15}
\end{figure}

\begin{figure}
\begin{center}
\resizebox{8cm}{!}{\includegraphics{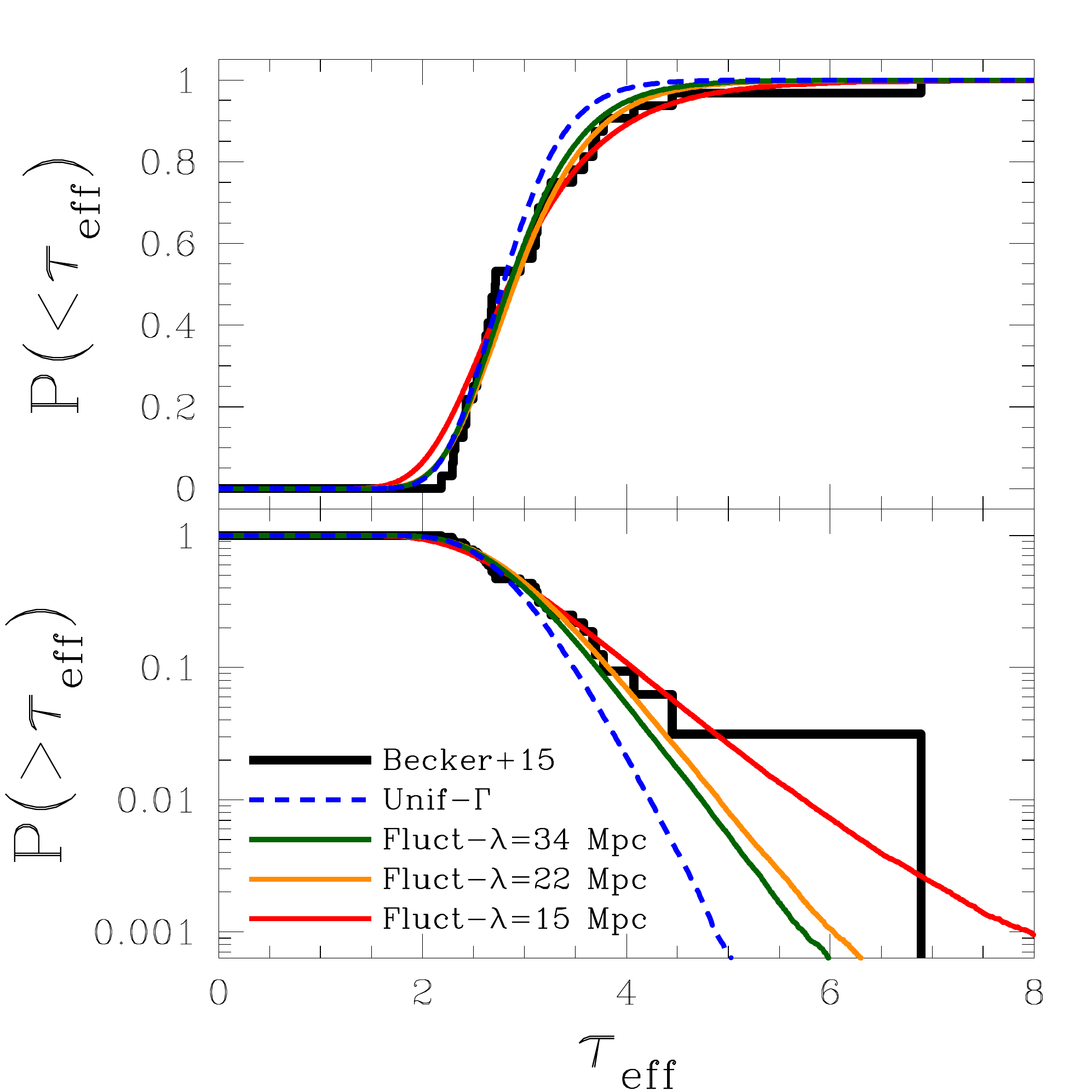}}\\
\end{center}
\caption{Top: Cumulative optical depth distribution $P(<\teff)$ computed for the uniform background (dashed blue curve) and fluctuating background models (solid color curves). The red, orange, and green curves show the fluctuating-$\lambda=15$, 22, and 34 Mpc models, respectively. The observed distribution from \citet{Becker2015}, including lower limits, is shown as the black curve. Bottom: The same curves as the top panel but recast as $P(>\teff)=1-P(<\teff)$ and shown on a logarithmic scale to emphasize the high-$\teff$ behavior.}
\label{fig:taudist}
\end{figure}

\begin{figure}
\begin{center}
\resizebox{8cm}{!}{\includegraphics{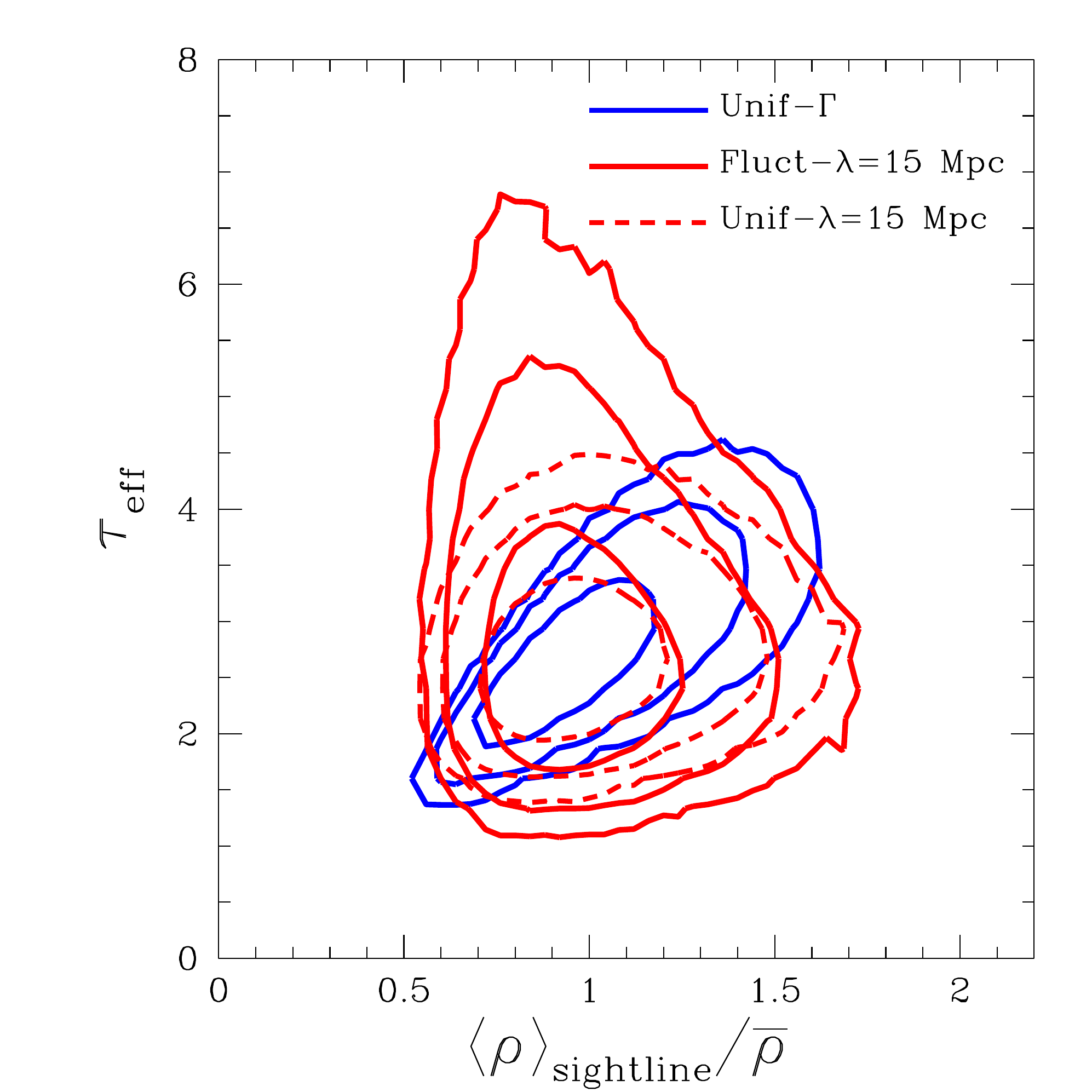}}\\
\end{center}
\caption{Relationship between average density along simulated sightlines and the resulting effective optical depth in the uniform (purple) and fluctuating (solid: fluctuating-$\lambda$, dashed: uniform-$\lambda$) ionizing background models. The contour levels enclose 68$\%$, 95$\%$, and 99$\%$ of sightlines. When the ionizing background is uniform, the large-scale opacity of the IGM is a tracer of the density field, as shown by the tight correlation of the blue contours (see also Figure~\ref{fig:lyamaps}). In contrast, in our fluctuating ionizing background model the most opaque regions of the \lya forest correspond to underdense regions where the radiation field is suppressed. The addition of mean free path fluctuations greatly increases this effect.}
\label{fig:contour}
\end{figure}

The uniform $\GHI$ slice in the left panel of Figure~\ref{fig:lyamaps} is simply a reflection of the projected density field -- regions with higher density show less transmission, and vice versa. This is the standard picture of the \lya forest at lower redshifts when the ionizing background is largely uniform. High density regions only make up a small portion of the volume in the standard cosmological model, so opaque sightlines are rare. Once the correlation between the density field and the radiation field is strong enough, as in the fluctuating-$\lambda$ model in the right panel of Figure~\ref{fig:lyamaps}, this picture reverses: low density regions become opaque due to a dearth of ionizing photons, at least on scales larger than the average mean free path.

To determine the statistical properties of large-scale features in the \lya forest throughout our simulation volume, we computed the \lya effective optical depth along 250000 randomly positioned and oriented 50 Mpc$/h$ sightlines in our fiducial realization. In Figure~\ref{fig:taudist15}, we compare the observations by B15 to our $\lambda=15$ Mpc models with and without mean free path fluctuations. The uniform background and uniform-$\lambda$ models are nearly identical, despite the clear difference in physical environments corresponding to a given optical depth seen in Figure~\ref{fig:lyamaps}. In agreement with B15 we find that a uniform ionizing background is sufficient to describe the observed $P(<\teff)<0.5$, but it fails to explain the tail to higher optical depths. Fluctuations in the mean free path naturally lead to a tail of high optical depths that are more consistent with observations than models that assume a uniform mean free path. 

In Figure~\ref{fig:taudist} we show the distribution of $\teff$ for $\lambda=$ 15, 22, and 34 Mpc fluctuating-$\lambda$ models. The highest $\teff$ observed by B15 is extremely rare unless the average $\lambda$ is small, and it is still quite rare even in our $\lambda=15$ Mpc model (but see Section~\ref{sec:cv}).  We also note that the $\lambda=15$ Mpc model shows some subtle deviations from observations at small $\teff$. These deviations come from regions experiencing a strong radiation field, i.e. environments close to galaxies, which our simplified density field model is unlikely to model accurately. For this reason, our model cannot be rigorously and quantitatively compared to the observations. Such an effort would require an improved treatment of the small-scale environments of the ionizing sources, in addition to our large-volume treatment that matches the observed tail to high $\teff$.

In Figure~\ref{fig:contour} we compare the relationship between the average density along each sightline and the resulting optical depth for the uniform ionizing background, uniform-$\lambda$, and fluctuating-$\lambda$ models. While the average density and optical depth are tightly correlated when the ionizing background is uniform, the situation becomes more complex when the radiation field strongly fluctuates. Although the distribution of $\teff$ is very similar between the uniform background and uniform-$\lambda$ cases (Figure~\ref{fig:taudist15}), the physical nature of a sightline with a given $\teff$ is different: in the uniform-$\lambda$ model the rare sightlines with high optical depth are modestly underdense and the densest regions are typically close to the median optical depth. This difference becomes greatly enhanced once mean free path fluctuations are included.

We note that the distribution of $\teff$ in our uniform background model (i.e. the left panel of Figure~\ref{fig:lyamaps} and the solid blue curve in Figure~\ref{fig:taudist}) is somewhat broader than in previous works \citep{Becker2015,Chardin2015}, likely due to our poor physical resolution and relatively large cosmological volume (see the Appendix of \citealt{Becker2015}). Note, however, that the broader distribution seen in our simulations makes our estimate of the impact of a fluctuating mean free path a conservative one. Recall that the introduction of a mean free path inverts the relation between density and optical depth, which is a positive correlation in the spatially constant $\Gamma$ case (i.e., the blue contours in Figure~\ref{fig:contour}). If that relation is \emph{weaker} for a spatially constant ionizing background, the introduction of the mean free path will result in even stronger fluctuations.  Testing this in detail is beyond the scope of this paper, as it requires numerical simulations with extremely high dynamic range.

\section{Evolution of the effective optical depth distribution from $z\sim5.8$--$5.4$}\label{sec:zevol}

In the preceding sections, we attempted to explain observations of the variations in \lya forest opacity at $z\sim5.6$ in the context of a radiation field regulated by a fluctuating mean free path. However, another interesting feature of the \lya forest at these redshifts is the rapid decline in the average transmission above $z\sim5.5$ \citep{Fan2006} coincident with increased variations in transmission between sightlines (B15). Here we examine whether the fluctuating mean free path model described above can account for this evolution. In particular, we consider models at $z=5.4$ and $z=5.8$ to compare with the $z=5.3$--$5.5$ and $z=5.7$--$5.9$ $\teff$ distributions from B15.

To construct models at different redshifts that are consistent with each other, we follow the same procedure from Section 2 to create density and halo fields at $z=5.4$ and $z=5.8$ using the same initial conditions. Keeping the ratio between ionizing and non-ionizing UV luminosity fixed, and for a fixed limiting UV magnitude (which is roughly consistent with our halo matching procedure), integration of the \citet{Bouwens2015b} luminosity functions (interpolated between $z\sim5$--6) results in an evolution of the ionizing emissivity $\epsilon_\mathrm{ion}\propto(1+z)^{-2}$. As a rough approximation for the evolution of $\lambda$ at fixed $\GHI$, we assume the power law evolution $\lambda\propto(1+z)^{-4.4}$ (comoving) measured by \citet{Worseck2014} across $2 < z < 5$ where $\GHI$ is roughly constant \citep{BB2013}.

\begin{figure}
\begin{center}
\resizebox{8cm}{!}{\includegraphics{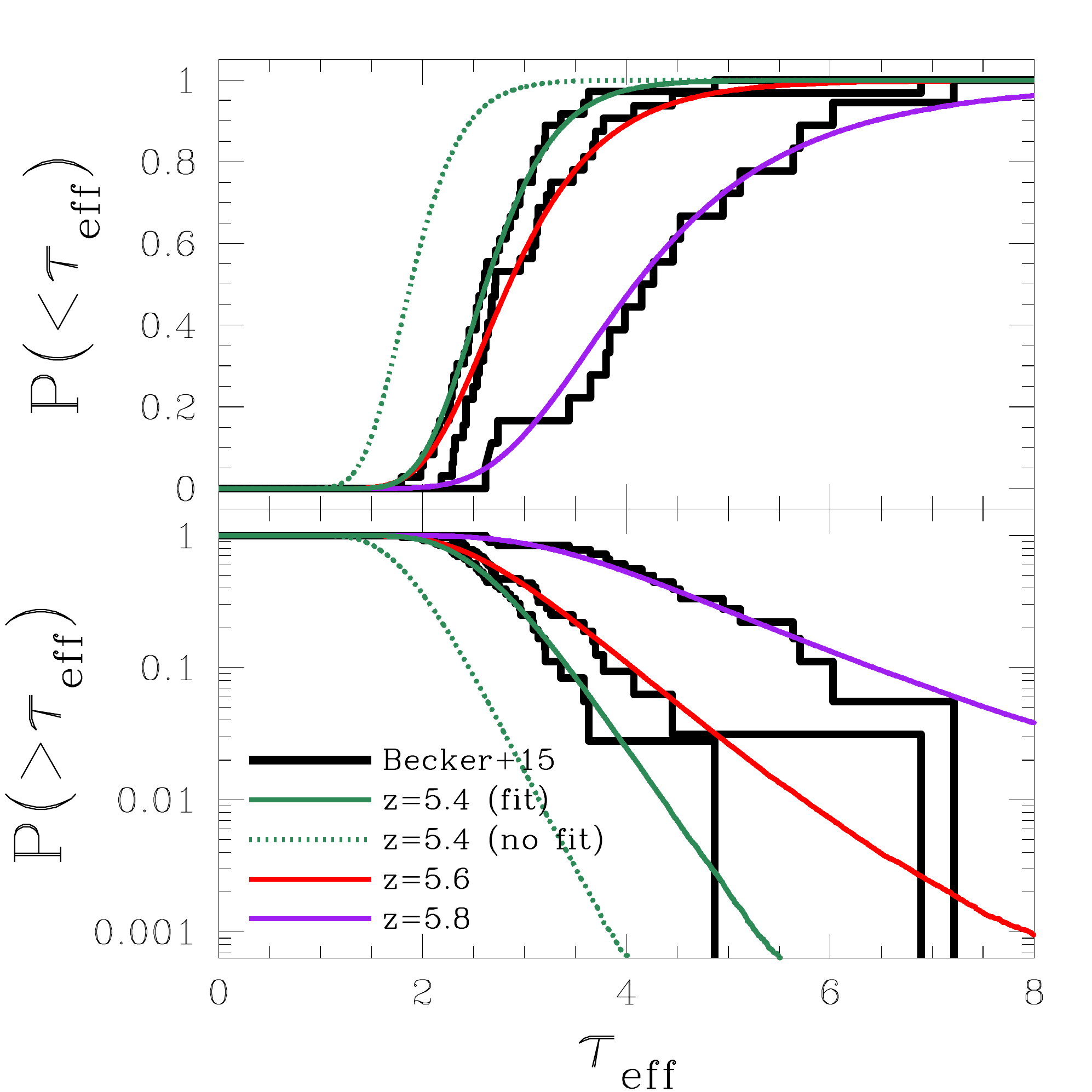}}\\
\end{center}
\caption{Top: Cumulative optical depth distributions $P(<\teff)$ in the fluctuating $\lambda$ model (green, red, purple curves) compared to the observations from \citet{Becker2015} (black curves) at $z\sim5.4, 5.6, 5.8$ from left to right. The dotted green curve shows the distribution at $z=5.4$ without renormalizing the optical depths. Bottom: The same curves as above, but recast as $P(>\teff)$ and shown on a logarithmic scale to emphasize the high-$\teff$ behavior.}
\label{fig:taudistevol}
\end{figure}

Primarily as a result of the rapid evolution in $\lambda$, in this model the average $\GHI$ increases by more than a factor of two from $z=5.8$ to $z=5.4$. In Figure~\ref{fig:taudistevol}, we show the resulting evolution of the cumulative $\teff$ distributions in our $\lambda=15$ Mpc model. The solid curves show the ``best-fit" distributions obtained by tuning the $\kappa$ parameter (equation~\ref{eqn:taugp}) separately at each redshift. The $z=5.6$ and $z=5.8$ curves require the same $\kappa$ to fit the observations, while $z=5.4$ requires a substantially higher value. The dotted curve shows the $z=5.4$ distribution using the same $\kappa$ as the other redshifts, demonstrating a clear disagreement with the measured values. While the rapid $\GHI$ evolution in our model is consistent with the observed evolution in the \lya forest transmission from $z\sim5.6$--$5.8$, it significantly underestimates $\teff$ at $z=5.4$, requiring a substantial adjustment in $\kappa$ from $\sim0.14$ to $\sim0.22$ to match observations. The simplest interpretation of this $\kappa$ adjustment is that our model overestimates $\GHI$ at $z=5.4$ by about a factor of $\sim1.6$, essentially requiring $\GHI$ to be constant from $z=5.6$--5.4 while increasing sharply from $z=5.8$--5.6. 

\section{Discussion}\label{sec:hidiscuss}

Models of the ionizing background that assume either a uniform background or a uniform mean free path of ionizing photons are unable to reproduce the distribution of Gunn-Peterson troughs at $z\ga5.4$ (B15). Using our semi-numerical ionizing background model, we find that introducing fluctuations in the mean free path greatly enhances the fluctuations in the strength of the ionizing background. The increase in fluctuations manifests on the large scales ($>\lambda$) required by observations. Mean free path fluctuations come about due to the inherently fluctuating ionizing background in the Universe immediately following reionization -- the average mean free path is short and early halos are strongly biased, so fluctuations in the ionizing background of roughly a factor of two above and below the mean are inevitable \citep{MF2009} with underdense regions experiencing a weaker ionizing background than overdense regions.

The resulting mean free path fluctuations then depend on the competition between two effects: the regulation of absorbers by the ionizing background (M11) and the relative number density of absorbers due to the large-scale density field. Under reasonable assumptions ($\lambda \propto \GHI^{2/3} \Delta^{-1}$), we find that on average the mean free path decreases in voids and increases in biased regions, leading to enhanced and diminished ionizing background strengths, respectively. Because the radiation field is effectively ``smoothed" by the mean free path, the effect of mean free path fluctuations is most prominent on scales larger than the average mean free path in the IGM.

Applying these enhanced ionizing background fluctuations back onto the simulated density field, we find that the typical picture of the \lya forest where transmission and density are anti-correlated reverses on large-scales (Figures~\ref{fig:lyamaps} and \ref{fig:contour}). The resulting distribution of effective optical depths is very similar to observations if the average mean free path is relatively short ($\lambda\la20$ Mpc), although in detail we find that our model has difficulty matching the low-$\teff$ and high-$\teff$ ends of the distribution simultaneously.

\subsection{Comparison to previous work}

The impact of mean free path fluctuations on the \lya forest has also been examined by \citet{Pontzen2014} and \citet{Gontcho2014}, albeit in the context of baryon acoustic oscillations measurements at $z\sim2.3$. \citet{Pontzen2014} showed that when mean free path fluctuations were included the bias of neutral hydrogen on scales larger than the mean free path becomes negative (see their Figure 2), in qualitative agreement with our results. In contrast to our non-linear 3D approach, both of these authors applied linear theory arguments to analytically compute the effect of mean free path fluctuations through scale-dependent bias of the radiation field. They found that the measured flux power spectrum contains valuable information that may allow constraints on the properties of sources of ionizing photons and the mean free path (see also \citealt{Pontzen2014b}). We leave simulations of the flux power spectrum to future work, but note that such arguments may still apply at $z>5$ albeit on much smaller scales.

As discussed above, our model provides a much better fit to observations of the \lya forest at $z\sim5.6$ than the uniform mean free path models in B15. However, a direct comparison is complicated by our different methods for simulating the density field, the B15 model having higher resolution and a much more accurate simulation of the gas physics. In contrast to B15, we do not find that uniform mean free path models of the fluctuating ionizing background produce a wider distribution of $\teff$ than uniform background models (see Figure~\ref{fig:taudist15}), but this may be due to our inclusion of much fainter sources ($M_\mathrm{UV}\la-13$ vs. $M_\mathrm{UV}\la-18$ in B15) which act to smooth the radiation field because of their weaker level of clustering.

The ionizing emissivity in our fiducial $\lambda=15$ Mpc model is $\epsilon_\mathrm{ion}\sim1.8\times10^{25} (\langle\GHI\rangle/10^{-12.5}$ s$^{-1})$ erg s$^{-1}$ Mpc$^{-3}$ Hz$^{-1}$. This compares favorably with the estimates by \citet{BB2013} and the required slow evolution to $z>6$ implied by reionization constraints \citep{Robertson2015,Bouwens2015a}. We cannot claim to constrain the true value of $\epsilon_\mathrm{ion}$ with our model --  in the context of our \lya forest modeling, the low $\kappa$ value we use implies a higher $\GHI$ by a factor of $\sim7$ (as compared to a naive extrapolation between the measurements at $z\sim5$ and $z\sim6$, e.g. \citealt{Calverley2011,WB2011}), but this ignores the substantial error in the computed transmission due to the low resolution of our simulated density field \citep{BB2009} and other uncertainties due to the thermal history of the IGM and inaccuracies inherent to the FGPA method (e.g. \citealt{Bolton2005}). In addition, the average mean free path required to reproduce the observed tail to high optical depths may differ depending on the $\Gamma$ and $\Delta$ dependencies of the mean free path. However, it is reassuring that our model is not grossly inconsistent with the current understanding of ionizing photon output at $z\sim5.5$.

Recently, \citet{Chardin2015} attempted to address the broad distribution of effective optical depths at $z>5$ by computing the radiative transfer of ionizing photons through small volume (10--40 Mpc on a side) high-resolution cosmological simulations. After the end of reionization, the ionizing background in their simulation volumes is extremely uniform, resulting in only minor variations in $\teff$ between sightlines. As a potential explanation for the broad distribution observed by B15 they presented a toy ``rare source" model where the ionizing background was dominated by a population of luminous quasars. Such a model is unlikely given the constraints on the number density of such sources (e.g. \citealt{McGreer2013}), and we suggest that a much simpler explanation for the uniformity of their simulated ionizing background exists. Because their simulation volumes are comparable or smaller than the mean free path after reionization (10--40 Mpc), the ``smoothing" effect of the mean free path on the radiation field naturally leads to an almost completely uniform background. We believe that high-resolution radiative transfer modeling, similar to that of \citet{Chardin2015}, is critical to understanding the regulation of neutral gas in the IGM by the ionizing background during and after reionization (e.g. \citealt{Rahmati2013}), but modeling the effect of mean free path fluctuations (i.e. large-scale coherent variations in the properties of absorbers) requires volumes that are at least several $\lambda$ on a side.

\subsection{Cosmic variance}\label{sec:cv}

\begin{figure}
\begin{center}
\resizebox{8cm}{!}{\includegraphics{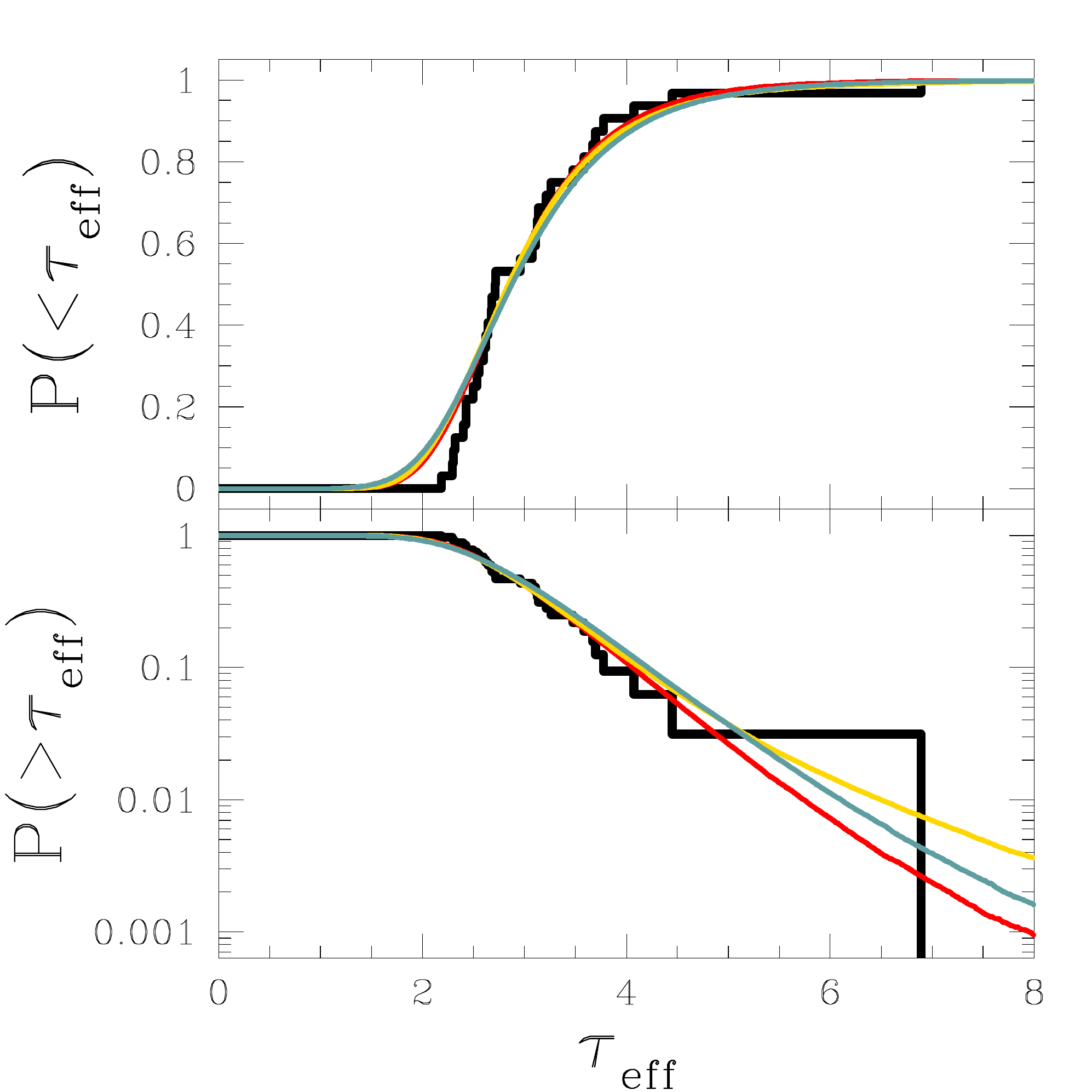}}\\
\end{center}
\caption{Top: Comparison of cumulative optical depth distributions $P(<\tau)$ for three different realizations of cosmological initial conditions (indicated by colors -- red is the fiducial model discussed in previous sections) with fluctuating-$\lambda=15$ Mpc radiation fields. In this panel, the three realizations are nearly identical. Bottom: The same curves as above, but recast as $P(>\teff)$ and shown on a logarithmic scale to emphasize the high-$\teff$ behavior. The different realizations are clearly discrepant at the high-$\teff$ end, indicating that the (400 Mpc)$^3$ volume still suffers from cosmic variance at the factor of $\sim3$ level at $\teff\sim7$.}
\label{fig:cv}
\end{figure}

The volume of the simulations presented in the preceding sections is 400 Mpc on a side, corresponding to $\ga10$ mean free paths. While this is considerably larger than previous computations of the fluctuating ionizing background (e.g. \citealt{MF2009,Becker2015}), it is worth examining whether this volume is in fact large enough to fully capture the fluctuations seen in observations. As a simple test, we ran two additional simulations with the same model parameters but different realizations of the cosmological initial conditions. Figure~\ref{fig:cv} shows the resulting $\teff$ distributions with fluctuating-$\lambda=15$ Mpc radiation fields at $z=5.6$ for all three realizations. The $\teff$ distributions are in agreement for $\teff\la4$, but they begin to diverge in the high-$\teff$ tail reaching a factor of $\sim3$ discrepancy at $\teff\sim7$. Note in particular that our fiducial model has the weakest tail at high $\teff$, with the other realizations slightly closer to the B15 observations at the highest optical depths.

The importance of cosmic variance in the (rare) high-$\teff$ tail is not surprising. In our model, long wavelength modes of the density field -- reflected in the distribution of sources and the resulting radiation field -- are the source of \lya forest opacity variations. However, there are only $\sim(400$ Mpc$)^3/(50$ Mpc$/h)^3\sim200$ independent regions in our simulated volume at the scale of the $\teff$ measurements, limiting the predictive power of each realization to features more common than the few-percent level. This argument ignores the fact that even larger modes also play an important role in producing coherent background fluctuations (see, e.g., the $\sim200\times50$ Mpc void in Figure~\ref{fig:denshalouvb}) which may require yet larger volumes to model accurately, beyond the reach of our existing computational resources. Such large features in the radiation field may require time-dependent modeling of ionizing source and IGM opacity evolution due to their substantial light-crossing time, an effect which we ignore in this work.

\subsection{Variation of ionizing source and absorber parameters}

The important assumptions we make for sources of ionizing photons in our model are as follows:
\begin{enumerate}
\item Duty cycle of unity
\item Fixed ratio of ionizing to non-ionizing UV luminosity
\item Minimum halo mass of $\sim2\times10^9 M_\odot$
\end{enumerate}
These three assumptions all relate to the effective bias of ionizing emissivity in our simulations. If the emissivity bias is larger, then the resulting fluctuations in the ionizing background and mean free path will be stronger. Decreasing the duty cycle from unity would result in an increased emissivity bias and thus result in stronger fluctuations, although likely only modestly (see: \citealt{MF2009}). The ratio of ionizing to non-ionizing UV luminosity is typically assumed to be fixed, but the stellar properties and environments of UV-faint and UV-luminous galaxies are likely to be different. Recent simulations by \citet{Wise2014} found that more massive (and thus more luminous) galaxies exhibited smaller $f_\mathrm{esc}$ than lower mass galaxies due to the relative robustness of their interstellar medium to supernova explosions. Reducing the ionizing photon output from high-luminosity galaxies would reduce the emissivity bias, leading to weaker fluctuations. However, due to the steep faint-end slope of the \citet{Bouwens2015b} UV luminosity function, the dominant contributors to the ionizing photon budget are already relatively faint galaxies ($M_\mathrm{UV}\sim-18$), so the effect on our results would likely be modest. For a similar reason, reducing the halo mass cutoff is unlikely to have a substantial effect on ionizing background fluctuations because the contribution from $M_\mathrm{UV}>-13$ galaxies to the UV luminosity density is small.

As discussed above, our mean free path model has two key parameters: the dependencies on $\GHI$ and $\Delta$. The mean free path should change with the strength of the ionizing background through regulation of the size of neutral absorbing clouds. Our assumption of $\lambda\propto\GHI^{2/3}$ is equivalent to assuming a gas density PDF $P(\Delta)\propto\Delta^{-2.5}$. This assumption may be somewhat conservative -- M11 found a steeper PDF slope at $z\sim5.6$ in their hydrodynamic cosmological simulations corresponding to $\lambda\propto\GHI^{\sim3/4}$. By running additional simulations with this steeper dependence, we find that it has a modest effect that is barely distinguishable from simply adjusting the average mean free path. For example, a $\lambda=$ 22 Mpc model run with $\lambda\propto\GHI^{3/4}$ is nearly identical to our fiducial model with $\lambda=$ 15 Mpc and $\lambda\propto\GHI^{2/3}$. 

The mean free path should also depend on the density of the local environment -- higher density regions will have more dense gas to absorb ionizing photons. For simplicity, we assumed that the number density of absorbers in a given volume is proportional to $\Delta$, or $\lambda\propto\Delta^{-1}$. The extremely flat evolution of the ionizing background at $z\sim2$--$5$ suggests that neutral gas is associated with dark matter halos \citep{Munoz2015} so we may be underestimating the effect that the density field has on the mean free path. However, at $z\sim6$, when the ionizing background is evolving extremely rapidly, the overdensities associated with self-shielded gas become smaller than the virial overdensity \citep{Munoz2015}, and so the bias of absorbers may be small. The sharp change in the inferred redshift evolution of the ionizing background at $z\sim5.5$ (e.g. \citealt{Fan2006}) may reflect a transition from neutral gas residing predominantly in the IGM to neutral gas residing predominantly in collapsed objects.

\section{Conclusion}\label{sec:hiconclude}

In this work, we have constructed a 3D semi-numerical model of the ionizing background that, for the first time, self-consistently includes the effect of fluctuations in the mean free path due to the ionizing radiation field and density field. We combine the semi-numerical halo model of {\small DEXM} (MF07) with a prescription for mean free path variations motivated by analytic calculations and hydrodynamic simulations (M11) and compute the ionizing background in a volume 400 Mpc on a side, iterating until the spatially variable mean free path and ionizing background are self-consistent. The resulting radiation field shows strongly enhanced fluctuations relative to previous models that assume a uniform mean free path. Applying the FGPA to sightlines through the quasi-linear density field in our simulation volume, we find that the addition of mean free path fluctuations substantially broadens the distribution of \lya forest effective optical depths on large scales, particularly at the high-$\teff$ end which eluded previous work.

Our results show that the end phases of reionization are rich in complexity and astrophysics: there is not a sharp transition from the epoch of reionization to the simple uniform background that is consistent with the \lya forest at lower redshift. Rather, the inhomogeneous source \emph{and} IGM distributions leave observable artifacts even after reionization is over. In principle, we can use these signatures to learn about the absorber population at the tail end of reionization and better understand how they regulate the final phases of that process. If our understanding of the IGM improves, this may allow us to model the disappearance of the last neutral regions in the Universe. The ``self-shielding" of void regions by absorbers at their edges (e.g. \citealt{Crociani2011}) may prolong the end stages of reionization by acting as large-scale sinks of ionizing photons, similar to the semi-analytic modeling by \citet{SM2014} which showed the importance of a $\Gamma$-dependent recombination rate.

Similar effects may be present in the \HeIIa \lya forest after \HeIIa reionization at $z\sim3$. Indeed, large variations in $\teff$ are seen on $\sim40$ Mpc scales down to $z\sim2.7$, and regions with relatively low $\teff$ appear to exist as early as $z\sim3.5$ \citep{Worseck2014a}. These observations have been interpreted as an early and extended \HeIIa reionization process \citep{Compostella2014}. In contrast to the hydrogen ionizing background model in this work, the dominant source of \HeIIa ionizing background fluctuations is the rarity of bright quasars that likely dominate the \HeIIb-ionizing photon budget (e.g. \citealt{Furlanetto2009}). These fluctuations may similarly ``seed" fluctuations in the mean free path, leading to large-scale variations in $\teff$ that mimic the end of reionization (Davies, Furlanetto, \& Dixon, in prep.).

Many avenues exist to improve the predictive power of our model. Potential modifications include a larger simulation volume to better characterize the effect of cosmic variance and long wavelength modes in the density field, high-resolution N-body and/or hydrodynamic simulations for a more realistic IGM, and more sophisticated models for variations in the mean free path as a function of environment. Understanding the radiative feedback between ionizing sources and neutral absorbers may be critical to explaining the sudden shift in \lya forest evolution at $z\sim5.5$ \citep{Fan2006,Becker2015,Munoz2015}, and the distribution of effective optical depths is a key piece of that puzzle.

In the final stages of preparing this paper, we became aware of \citet{Daloisio2015}, which presents an alternative explanation for the broad distribution of $\teff$ at $z>5$. In their model an extended reionization process, ending at $z\sim6$, leads to strong spatial variations in residual heat, which lead to corresponding large-scale variations in $\teff$ ($\tau_\mathrm{GP}\propto T^{-0.724}$, see equation~\ref{eqn:taugp}). The resulting relationship between overdensity and $\teff$ is reversed compared to our model: voids, which are ionized later, remain hot (i.e. transparent), while dense regions, which are ionized first, have had time to cool and become more opaque (although this relationship is not taken into account in their transmission spectra). This effect should in principle dilute the effect of a fluctuating ionizing background, and it remains to be seen which effect (if either) dominates.

Because of the direct connection between large-scale features in the galaxy population to Gunn-Peterson troughs in the IGM, it may be possible to directly test our model through surveys of $z\sim5.5$--$5.9$ galaxies in $z\ga6$ QSO fields. For example, comparing the distribution of galaxies in the upper right panel of Figure~\ref{fig:denshalouvb} to the effective optical depth map in Figure~\ref{fig:lyamaps}, the number of $\ga{L}_*$ galaxies in fields corresponding to sightlines with large opaque troughs at the same redshift is much smaller than fields with excess transmission. In future work we will use our model to make predictions for the correlation between large-scale \lya forest $\teff$ and UV-selected galaxy populations that can be directly tested with current ground-based observatories. In principle these predictions would allow a definitive test of our model versus the \citet{Daloisio2015} model. 

\section*{Acknowledgements}

We would like to thank Alice Shapley, George Becker, James Bolton, Matthew McQuinn, and Anson D'Aloisio for helpful comments on a draft of this manuscript. We would also like to thank Joseph Hennawi and the ENIGMA group at MPIA for additional comments and suggestions which greatly improved the final draft. SRF was partially supported by NASA grant ATP-NNX15AK80G and by the Simons Foundation.

\bibliographystyle{mnras}
\bibliography{refs}

\begin{thebibliography}{}
\makeatletter
\relax
\def\mn@urlcharsother{\let\do\@makeother \do\$\do\&\do\#\do\^\do\_\do\%\do\~}
\def\mn@doi{\begingroup\mn@urlcharsother \@ifnextchar [ {\mn@doi@}
  {\mn@doi@[]}}
\def\mn@doi@[#1]#2{\def\@tempa{#1}\ifx\@tempa\@empty \href
  {http://dx.doi.org/#2} {doi:#2}\else \href {http://dx.doi.org/#2} {#1}\fi
  \endgroup}
\def\mn@eprint#1#2{\mn@eprint@#1:#2::\@nil}
\def\mn@eprint@arXiv#1{\href {http://arxiv.org/abs/#1} {{\tt arXiv:#1}}}
\def\mn@eprint@dblp#1{\href {http://dblp.uni-trier.de/rec/bibtex/#1.xml}
  {dblp:#1}}
\def\mn@eprint@#1:#2:#3:#4\@nil{\def\@tempa {#1}\def\@tempb {#2}\def\@tempc
  {#3}\ifx \@tempc \@empty \let \@tempc \@tempb \let \@tempb \@tempa \fi \ifx
  \@tempb \@empty \def\@tempb {arXiv}\fi \@ifundefined
  {mn@eprint@\@tempb}{\@tempb:\@tempc}{\expandafter \expandafter \csname
  mn@eprint@\@tempb\endcsname \expandafter{\@tempc}}}

\bibitem[\protect\citeauthoryear{Becker \& Bolton}{Becker \&
  Bolton}{2013}]{BB2013}
Becker G.~D.,  Bolton J.~S.,  2013, MNRAS, 436, 1023

\bibitem[\protect\citeauthoryear{Becker, Bolton, Madau, Pettini, Ryan-Weber  \&
  Venemans}{Becker et~al.}{2015}]{Becker2015}
Becker G.~D.,  Bolton J.~S.,  Madau P.,  Pettini M.,  Ryan-Weber E.~V.,
  Venemans B.~P.,  2015, MNRAS, 447, 3402

\bibitem[\protect\citeauthoryear{{Bolton} \& {Becker}}{{Bolton} \&
  {Becker}}{2009}]{BB2009}
{Bolton} J.~S.,  {Becker} G.~D.,  2009, \mn@doi [\mnras]
  {10.1111/j.1745-3933.2009.00700.x}, \href
  {http://adsabs.harvard.edu/abs/2009MNRAS.398L..26B} {398, L26}

\bibitem[\protect\citeauthoryear{{Bolton} \& {Haehnelt}}{{Bolton} \&
  {Haehnelt}}{2007}]{BH2007a}
{Bolton} J.~S.,  {Haehnelt} M.~G.,  2007, \mn@doi [\mnras]
  {10.1111/j.1365-2966.2007.12372.x}, \href
  {http://adsabs.harvard.edu/abs/2007MNRAS.382..325B} {382, 325}

\bibitem[\protect\citeauthoryear{{Bolton} \& {Haehnelt}}{{Bolton} \&
  {Haehnelt}}{2013}]{BH2013}
{Bolton} J.~S.,  {Haehnelt} M.~G.,  2013, \mn@doi [\mnras]
  {10.1093/mnras/sts455}, \href
  {http://adsabs.harvard.edu/abs/2013MNRAS.429.1695B} {429, 1695}

\bibitem[\protect\citeauthoryear{{Bolton}, {Haehnelt}, {Viel}  \&
  {Springel}}{{Bolton} et~al.}{2005}]{Bolton2005}
{Bolton} J.~S.,  {Haehnelt} M.~G.,  {Viel} M.,   {Springel} V.,  2005, \mn@doi
  [\mnras] {10.1111/j.1365-2966.2005.08704.x}, \href
  {http://adsabs.harvard.edu/abs/2005MNRAS.357.1178B} {357, 1178}

\bibitem[\protect\citeauthoryear{{Bolton}, {Haehnelt}, {Warren}, {Hewett},
  {Mortlock}, {Venemans}, {McMahon}  \& {Simpson}}{{Bolton}
  et~al.}{2011}]{Bolton2011}
{Bolton} J.~S.,  {Haehnelt} M.~G.,  {Warren} S.~J.,  {Hewett} P.~C.,
  {Mortlock} D.~J.,  {Venemans} B.~P.,  {McMahon} R.~G.,   {Simpson} C.,  2011,
  \mn@doi [\mnras] {10.1111/j.1745-3933.2011.01100.x}, \href
  {http://adsabs.harvard.edu/abs/2011MNRAS.416L..70B} {416, L70}

\bibitem[\protect\citeauthoryear{{Bosman} \& {Becker}}{{Bosman} \&
  {Becker}}{2015}]{BB2015}
{Bosman} S.~E.~I.,  {Becker} G.~D.,  2015, preprint, \href
  {http://adsabs.harvard.edu/abs/2015arXiv150506880B} {} (\mn@eprint {arXiv}
  {1505.06880})

\bibitem[\protect\citeauthoryear{{Bouwens}, {Illingworth}, {Oesch}, {Caruana},
  {Holwerda}, {Smit}  \& {Wilkins}}{{Bouwens} et~al.}{2015a}]{Bouwens2015a}
{Bouwens} R.~J.,  {Illingworth} G.~D.,  {Oesch} P.~A.,  {Caruana} J.,
  {Holwerda} B.,  {Smit} R.,   {Wilkins} S.,  2015a, preprint, \href
  {http://adsabs.harvard.edu/abs/2015arXiv150308228B} {} (\mn@eprint {arXiv}
  {1503.08228})

\bibitem[\protect\citeauthoryear{{Bouwens} et~al.,}{{Bouwens}
  et~al.}{2015b}]{Bouwens2015b}
{Bouwens} R.~J.,  et~al., 2015b, \mn@doi [\apj] {10.1088/0004-637X/803/1/34},
  \href {http://adsabs.harvard.edu/abs/2015ApJ...803...34B} {803, 34}

\bibitem[\protect\citeauthoryear{{Calverley}, {Becker}, {Haehnelt}  \&
  {Bolton}}{{Calverley} et~al.}{2011}]{Calverley2011}
{Calverley} A.~P.,  {Becker} G.~D.,  {Haehnelt} M.~G.,   {Bolton} J.~S.,  2011,
  \mn@doi [\mnras] {10.1111/j.1365-2966.2010.18072.x}, \href
  {http://adsabs.harvard.edu/abs/2011MNRAS.412.2543C} {412, 2543}

\bibitem[\protect\citeauthoryear{{Chardin}, {Haehnelt}, {Aubert}  \&
  {Puchwein}}{{Chardin} et~al.}{2015}]{Chardin2015}
{Chardin} J.,  {Haehnelt} M.~G.,  {Aubert} D.,   {Puchwein} E.,  2015,
  preprint, \href {http://adsabs.harvard.edu/abs/2015arXiv150501853C} {}
  (\mn@eprint {arXiv} {1505.01853})

\bibitem[\protect\citeauthoryear{{Choudhury}, {Puchwein}, {Haehnelt}  \&
  {Bolton}}{{Choudhury} et~al.}{2014}]{Choudhury2014}
{Choudhury} T.~R.,  {Puchwein} E.,  {Haehnelt} M.~G.,   {Bolton} J.~S.,  2014,
  preprint, \href {http://adsabs.harvard.edu/abs/2014arXiv1412.4790C} {}
  (\mn@eprint {arXiv} {1412.4790})

\bibitem[\protect\citeauthoryear{Compostella, Cantalupo  \&
  Porciani}{Compostella et~al.}{2014}]{Compostella2014}
Compostella M.,  Cantalupo S.,   Porciani C.,  2014, MNRAS, 445, 4186

\bibitem[\protect\citeauthoryear{{Crociani}, {Mesinger}, {Moscardini}  \&
  {Furlanetto}}{{Crociani} et~al.}{2011}]{Crociani2011}
{Crociani} D.,  {Mesinger} A.,  {Moscardini} L.,   {Furlanetto} S.,  2011,
  \mn@doi [\mnras] {10.1111/j.1365-2966.2010.17680.x}, \href
  {http://adsabs.harvard.edu/abs/2011MNRAS.411..289C} {411, 289}

\bibitem[\protect\citeauthoryear{{D'Aloisio}, {McQuinn}  \& {Trac}}{{D'Aloisio}
  et~al.}{2015}]{Daloisio2015}
{D'Aloisio} A.,  {McQuinn} M.~J.,   {Trac} H.,  2015, preprint, \href
  {http://adsabs.harvard.edu/abs/2015arXiv150902523D} {} (\mn@eprint {arXiv}
  {1509.02523})

\bibitem[\protect\citeauthoryear{{Dijkstra}, {Mesinger}  \&
  {Wyithe}}{{Dijkstra} et~al.}{2011}]{Dijkstra2011}
{Dijkstra} M.,  {Mesinger} A.,   {Wyithe} J.~S.~B.,  2011, \mn@doi [\mnras]
  {10.1111/j.1365-2966.2011.18530.x}, \href
  {http://adsabs.harvard.edu/abs/2011MNRAS.414.2139D} {414, 2139}

\bibitem[\protect\citeauthoryear{{Dixon} \& {Furlanetto}}{{Dixon} \&
  {Furlanetto}}{2009}]{DF2009}
{Dixon} K.~L.,  {Furlanetto} S.~R.,  2009, \mn@doi [\apj]
  {10.1088/0004-637X/706/2/970}, \href
  {http://adsabs.harvard.edu/abs/2009ApJ...706..970D} {706, 970}

\bibitem[\protect\citeauthoryear{{Efstathiou}, {Davis}, {White}  \&
  {Frenk}}{{Efstathiou} et~al.}{1985}]{Efstathiou1985}
{Efstathiou} G.,  {Davis} M.,  {White} S.~D.~M.,   {Frenk} C.~S.,  1985,
  \mn@doi [\apjs] {10.1086/191003}, \href
  {http://adsabs.harvard.edu/abs/1985ApJS...57..241E} {57, 241}

\bibitem[\protect\citeauthoryear{{Fan} et~al.,}{{Fan} et~al.}{2001}]{Fan2001}
{Fan} X.,  et~al., 2001, \mn@doi [\aj] {10.1086/324111}, \href
  {http://adsabs.harvard.edu/abs/2001AJ....122.2833F} {122, 2833}

\bibitem[\protect\citeauthoryear{{Fan} et~al.,}{{Fan} et~al.}{2006}]{Fan2006}
{Fan} X.,  et~al., 2006, \mn@doi [\aj] {10.1086/504836}, \href
  {http://adsabs.harvard.edu/abs/2006AJ....132..117F} {132, 117}

\bibitem[\protect\citeauthoryear{Faucher-Gigu{\`e}re, Lidz, Zaldarriaga  \&
  Hernquist}{Faucher-Gigu{\`e}re et~al.}{2009}]{FG2009}
Faucher-Gigu{\`e}re C.-A.,  Lidz A.,  Zaldarriaga M.,   Hernquist L.,  2009,
  \mn@doi [ApJ] {10.1088/0004-637X/703/2/1416}, 703, 1416

\bibitem[\protect\citeauthoryear{Furlanetto}{Furlanetto}{2009}]{Furlanetto2009}
Furlanetto S.~R.,  2009, \mn@doi [ApJ] {10.1088/0004-637X/703/1/702}, 703, 702

\bibitem[\protect\citeauthoryear{{Furlanetto}, {Hernquist}  \&
  {Zaldarriaga}}{{Furlanetto} et~al.}{2004}]{Furlanetto2004}
{Furlanetto} S.~R.,  {Hernquist} L.,   {Zaldarriaga} M.,  2004, \mn@doi
  [\mnras] {10.1111/j.1365-2966.2004.08225.x}, \href
  {http://adsabs.harvard.edu/abs/2004MNRAS.354..695F} {354, 695}

\bibitem[\protect\citeauthoryear{{Gontcho A Gontcho}, {Miralda-Escud{\'e}}  \&
  {Busca}}{{Gontcho A Gontcho} et~al.}{2014}]{Gontcho2014}
{Gontcho A Gontcho} S.,  {Miralda-Escud{\'e}} J.,   {Busca} N.~G.,  2014,
  preprint, \href {http://adsabs.harvard.edu/abs/2014arXiv1404.7425G} {}
  (\mn@eprint {arXiv} {1404.7425})

\bibitem[\protect\citeauthoryear{{Gunn} \& {Peterson}}{{Gunn} \&
  {Peterson}}{1965}]{GP1965}
{Gunn} J.~E.,  {Peterson} B.~A.,  1965, \mn@doi [\apj] {10.1086/148444}, \href
  {http://adsabs.harvard.edu/abs/1965ApJ...142.1633G} {142, 1633}

\bibitem[\protect\citeauthoryear{Haardt \& Madau}{Haardt \&
  Madau}{1996}]{HM1996}
Haardt F.,  Madau P.,  1996, \mn@doi [ApJ] {10.1086/177035}, 461, 20

\bibitem[\protect\citeauthoryear{Haardt \& Madau}{Haardt \&
  Madau}{2012}]{HM2012}
Haardt F.,  Madau P.,  2012, \mn@doi [ApJ] {10.1088/0004-637X/746/2/125}, 746,
  125

\bibitem[\protect\citeauthoryear{{Hinshaw} et~al.,}{{Hinshaw}
  et~al.}{2013}]{Hinshaw2013}
{Hinshaw} G.,  et~al., 2013, \mn@doi [\apjs] {10.1088/0067-0049/208/2/19},
  \href {http://adsabs.harvard.edu/abs/2013ApJS..208...19H} {208, 19}

\bibitem[\protect\citeauthoryear{{Loeb} \& {Furlanetto}}{{Loeb} \&
  {Furlanetto}}{2013}]{stevebook}
{Loeb} A.,  {Furlanetto} S.~R.,  2013, {The First Galaxies in the Universe}.
{Princeton, NJ: Princeton University Press}

\bibitem[\protect\citeauthoryear{{McDonald}, {Seljak}, {Cen}, {Bode}  \&
  {Ostriker}}{{McDonald} et~al.}{2005}]{McDonald2005}
{McDonald} P.,  {Seljak} U.,  {Cen} R.,  {Bode} P.,   {Ostriker} J.~P.,  2005,
  \mn@doi [\mnras] {10.1111/j.1365-2966.2005.09141.x}, \href
  {http://adsabs.harvard.edu/abs/2005MNRAS.360.1471M} {360, 1471}

\bibitem[\protect\citeauthoryear{{McGreer}, {Mesinger}  \& {Fan}}{{McGreer}
  et~al.}{2011}]{McGreer2011}
{McGreer} I.~D.,  {Mesinger} A.,   {Fan} X.,  2011, \mn@doi [\mnras]
  {10.1111/j.1365-2966.2011.18935.x}, \href
  {http://adsabs.harvard.edu/abs/2011MNRAS.415.3237M} {415, 3237}

\bibitem[\protect\citeauthoryear{{McGreer} et~al.,}{{McGreer}
  et~al.}{2013}]{McGreer2013}
{McGreer} I.~D.,  et~al., 2013, \mn@doi [\apj] {10.1088/0004-637X/768/2/105},
  \href {http://adsabs.harvard.edu/abs/2013ApJ...768..105M} {768, 105}

\bibitem[\protect\citeauthoryear{{McGreer}, {Mesinger}  \&
  {D'Odorico}}{{McGreer} et~al.}{2015}]{McGreer2015}
{McGreer} I.~D.,  {Mesinger} A.,   {D'Odorico} V.,  2015, \mn@doi [\mnras]
  {10.1093/mnras/stu2449}, \href
  {http://adsabs.harvard.edu/abs/2015MNRAS.447..499M} {447, 499}

\bibitem[\protect\citeauthoryear{{McQuinn}, {Hernquist}, {Zaldarriaga}  \&
  {Dutta}}{{McQuinn} et~al.}{2007}]{McQuinn2007}
{McQuinn} M.,  {Hernquist} L.,  {Zaldarriaga} M.,   {Dutta} S.,  2007, \mn@doi
  [\mnras] {10.1111/j.1365-2966.2007.12085.x}, \href
  {http://adsabs.harvard.edu/abs/2007MNRAS.381...75M} {381, 75}

\bibitem[\protect\citeauthoryear{McQuinn, Oh  \& Faucher-Gigu{\`e}re}{McQuinn
  et~al.}{2011}]{McQuinn2011}
McQuinn M.,  Oh S.~P.,   Faucher-Gigu{\`e}re C.-A.,  2011, ApJ, 743, 82

\bibitem[\protect\citeauthoryear{Meiksin \& White}{Meiksin \&
  White}{2003}]{MW2003}
Meiksin A.,  White M.,  2003, \mn@doi [MNRAS]
  {10.1046/j.1365-8711.2003.06624.x}, 342, 1205

\bibitem[\protect\citeauthoryear{{Meiksin} \& {White}}{{Meiksin} \&
  {White}}{2004}]{MW2004}
{Meiksin} A.,  {White} M.,  2004, \mn@doi [\mnras]
  {10.1111/j.1365-2966.2004.07724.x}, \href
  {http://adsabs.harvard.edu/abs/2004MNRAS.350.1107M} {350, 1107}

\bibitem[\protect\citeauthoryear{{Mesinger}}{{Mesinger}}{2010}]{Mesinger2010}
{Mesinger} A.,  2010, \mn@doi [\mnras] {10.1111/j.1365-2966.2010.16995.x},
  \href {http://adsabs.harvard.edu/abs/2010MNRAS.407.1328M} {407, 1328}

\bibitem[\protect\citeauthoryear{{Mesinger} \& {Furlanetto}}{{Mesinger} \&
  {Furlanetto}}{2007}]{MF2007}
{Mesinger} A.,  {Furlanetto} S.,  2007, \mn@doi [\apj] {10.1086/521806}, \href
  {http://adsabs.harvard.edu/abs/2007ApJ...669..663M} {669, 663}

\bibitem[\protect\citeauthoryear{{Mesinger} \& {Furlanetto}}{{Mesinger} \&
  {Furlanetto}}{2008a}]{MF2008b}
{Mesinger} A.,  {Furlanetto} S.~R.,  2008a, \mn@doi [\mnras]
  {10.1111/j.1365-2966.2007.12836.x}, \href
  {http://adsabs.harvard.edu/abs/2008MNRAS.385.1348M} {385, 1348}

\bibitem[\protect\citeauthoryear{{Mesinger} \& {Furlanetto}}{{Mesinger} \&
  {Furlanetto}}{2008b}]{MF2008a}
{Mesinger} A.,  {Furlanetto} S.~R.,  2008b, \mn@doi [\mnras]
  {10.1111/j.1365-2966.2008.13039.x}, \href
  {http://adsabs.harvard.edu/abs/2008MNRAS.386.1990M} {386, 1990}

\bibitem[\protect\citeauthoryear{Mesinger \& Furlanetto}{Mesinger \&
  Furlanetto}{2009}]{MF2009}
Mesinger A.,  Furlanetto S.,  2009, MNRAS, 400, 1461

\bibitem[\protect\citeauthoryear{{Mesinger} \& {Haiman}}{{Mesinger} \&
  {Haiman}}{2007}]{MH2007}
{Mesinger} A.,  {Haiman} Z.,  2007, \mn@doi [\apj] {10.1086/513688}, \href
  {http://adsabs.harvard.edu/abs/2007ApJ...660..923M} {660, 923}

\bibitem[\protect\citeauthoryear{{Mesinger}, {Aykutalp}, {Vanzella},
  {Pentericci}, {Ferrara}  \& {Dijkstra}}{{Mesinger}
  et~al.}{2015}]{Mesinger2015}
{Mesinger} A.,  {Aykutalp} A.,  {Vanzella} E.,  {Pentericci} L.,  {Ferrara} A.,
    {Dijkstra} M.,  2015, \mn@doi [\mnras] {10.1093/mnras/stu2089}, \href
  {http://adsabs.harvard.edu/abs/2015MNRAS.446..566M} {446, 566}

\bibitem[\protect\citeauthoryear{{Miralda-Escud{\'e}}}{{Miralda-Escud{\'e}}}{1998}]{ME1998}
{Miralda-Escud{\'e}} J.,  1998, \mn@doi [\apj] {10.1086/305799}, \href
  {http://adsabs.harvard.edu/abs/1998ApJ...501...15M} {501, 15}

\bibitem[\protect\citeauthoryear{{Miralda-Escud{\'e}}, {Haehnelt}  \&
  {Rees}}{{Miralda-Escud{\'e}} et~al.}{2000}]{ME2000}
{Miralda-Escud{\'e}} J.,  {Haehnelt} M.,   {Rees} M.~J.,  2000, \mn@doi [\apj]
  {10.1086/308330}, \href {http://adsabs.harvard.edu/abs/2000ApJ...530....1M}
  {530, 1}

\bibitem[\protect\citeauthoryear{{Monaghan}}{{Monaghan}}{1992}]{Monaghan1992}
{Monaghan} J.~J.,  1992, \mn@doi [\araa] {10.1146/annurev.aa.30.090192.002551},
  \href {http://adsabs.harvard.edu/abs/1992ARA%26A..30..543M} {30, 543}

\bibitem[\protect\citeauthoryear{{Mu{\~n}oz}, {Oh}, {Davies}  \&
  {Furlanetto}}{{Mu{\~n}oz} et~al.}{2014}]{Munoz2015}
{Mu{\~n}oz} J.~A.,  {Oh} S.~P.,  {Davies} F.~B.,   {Furlanetto} S.~R.,  2014,
  preprint, \href {http://adsabs.harvard.edu/abs/2014arXiv1410.2249M} {}
  (\mn@eprint {arXiv} {1410.2249})

\bibitem[\protect\citeauthoryear{{Ono} et~al.,}{{Ono} et~al.}{2012}]{Ono2012}
{Ono} Y.,  et~al., 2012, \mn@doi [\apj] {10.1088/0004-637X/744/2/83}, \href
  {http://adsabs.harvard.edu/abs/2012ApJ...744...83O} {744, 83}

\bibitem[\protect\citeauthoryear{{Pentericci} et~al.,}{{Pentericci}
  et~al.}{2014}]{Pentericci2014}
{Pentericci} L.,  et~al., 2014, \mn@doi [\apj] {10.1088/0004-637X/793/2/113},
  \href {http://adsabs.harvard.edu/abs/2014ApJ...793..113P} {793, 113}

\bibitem[\protect\citeauthoryear{{Planck Collaboration} et~al.,}{{Planck
  Collaboration} et~al.}{2015}]{Planck2015}
{Planck Collaboration} et~al., 2015, preprint, \href
  {http://adsabs.harvard.edu/abs/2015arXiv150201589P} {} (\mn@eprint {arXiv}
  {1502.01589})

\bibitem[\protect\citeauthoryear{{Pontzen}}{{Pontzen}}{2014}]{Pontzen2014}
{Pontzen} A.,  2014, \mn@doi [\prd] {10.1103/PhysRevD.89.083010}, \href
  {http://adsabs.harvard.edu/abs/2014PhRvD..89h3010P} {89, 083010}

\bibitem[\protect\citeauthoryear{{Pontzen}, {Bird}, {Peiris}  \&
  {Verde}}{{Pontzen} et~al.}{2014}]{Pontzen2014b}
{Pontzen} A.,  {Bird} S.,  {Peiris} H.,   {Verde} L.,  2014, \mn@doi [\apjl]
  {10.1088/2041-8205/792/2/L34}, \href
  {http://adsabs.harvard.edu/abs/2014ApJ...792L..34P} {792, L34}

\bibitem[\protect\citeauthoryear{{Puchwein}, {Bolton}, {Haehnelt}, {Madau},
  {Becker}  \& {Haardt}}{{Puchwein} et~al.}{2015}]{Puchwein2015}
{Puchwein} E.,  {Bolton} J.~S.,  {Haehnelt} M.~G.,  {Madau} P.,  {Becker}
  G.~D.,   {Haardt} F.,  2015, \mn@doi [\mnras] {10.1093/mnras/stv773}, \href
  {http://adsabs.harvard.edu/abs/2015MNRAS.450.4081P} {450, 4081}

\bibitem[\protect\citeauthoryear{{Rahmati}, {Pawlik}, {Rai{\v c}evi{\` c}}  \&
  {Schaye}}{{Rahmati} et~al.}{2013}]{Rahmati2013}
{Rahmati} A.,  {Pawlik} A.~H.,  {Rai{\v c}evi{\` c}} M.,   {Schaye} J.,  2013,
  \mn@doi [\mnras] {10.1093/mnras/stt066}, \href
  {http://adsabs.harvard.edu/abs/2013MNRAS.430.2427R} {430, 2427}

\bibitem[\protect\citeauthoryear{{Robertson}, {Ellis}, {Furlanetto}  \&
  {Dunlop}}{{Robertson} et~al.}{2015}]{Robertson2015}
{Robertson} B.~E.,  {Ellis} R.~S.,  {Furlanetto} S.~R.,   {Dunlop} J.~S.,
  2015, \mn@doi [\apjl] {10.1088/2041-8205/802/2/L19}, \href
  {http://adsabs.harvard.edu/abs/2015ApJ...802L..19R} {802, L19}

\bibitem[\protect\citeauthoryear{{Rudie}, {Steidel}, {Shapley}  \&
  {Pettini}}{{Rudie} et~al.}{2013}]{Rudie2013}
{Rudie} G.~C.,  {Steidel} C.~C.,  {Shapley} A.~E.,   {Pettini} M.,  2013,
  \mn@doi [\apj] {10.1088/0004-637X/769/2/146}, \href
  {http://adsabs.harvard.edu/abs/2013ApJ...769..146R} {769, 146}

\bibitem[\protect\citeauthoryear{{Santos}}{{Santos}}{2004}]{Santos2004}
{Santos} M.~R.,  2004, \mn@doi [\mnras] {10.1111/j.1365-2966.2004.07594.x},
  \href {http://adsabs.harvard.edu/abs/2004MNRAS.349.1137S} {349, 1137}

\bibitem[\protect\citeauthoryear{{Schenker}, {Ellis}, {Konidaris}  \&
  {Stark}}{{Schenker} et~al.}{2014}]{Schenker2014}
{Schenker} M.~A.,  {Ellis} R.~S.,  {Konidaris} N.~P.,   {Stark} D.~P.,  2014,
  \mn@doi [\apj] {10.1088/0004-637X/795/1/20}, \href
  {http://adsabs.harvard.edu/abs/2014ApJ...795...20S} {795, 20}

\bibitem[\protect\citeauthoryear{{Schroeder}, {Mesinger}  \&
  {Haiman}}{{Schroeder} et~al.}{2013}]{Schroeder2013}
{Schroeder} J.,  {Mesinger} A.,   {Haiman} Z.,  2013, \mn@doi [\mnras]
  {10.1093/mnras/sts253}, \href
  {http://adsabs.harvard.edu/abs/2013MNRAS.428.3058S} {428, 3058}

\bibitem[\protect\citeauthoryear{{Sheth} \& {Tormen}}{{Sheth} \&
  {Tormen}}{1999}]{ST1999}
{Sheth} R.~K.,  {Tormen} G.,  1999, \mn@doi [\mnras]
  {10.1046/j.1365-8711.1999.02692.x}, \href
  {http://adsabs.harvard.edu/abs/1999MNRAS.308..119S} {308, 119}

\bibitem[\protect\citeauthoryear{{Simcoe}, {Sullivan}, {Cooksey}, {Kao},
  {Matejek}  \& {Burgasser}}{{Simcoe} et~al.}{2012}]{Simcoe2012}
{Simcoe} R.~A.,  {Sullivan} P.~W.,  {Cooksey} K.~L.,  {Kao} M.~M.,  {Matejek}
  M.~S.,   {Burgasser} A.~J.,  2012, \mn@doi [\nat] {10.1038/nature11612},
  \href {http://adsabs.harvard.edu/abs/2012Natur.492...79S} {492, 79}

\bibitem[\protect\citeauthoryear{{Sobacchi} \& {Mesinger}}{{Sobacchi} \&
  {Mesinger}}{2014}]{SM2014}
{Sobacchi} E.,  {Mesinger} A.,  2014, \mn@doi [\mnras] {10.1093/mnras/stu377},
  \href {http://adsabs.harvard.edu/abs/2014MNRAS.440.1662S} {440, 1662}

\bibitem[\protect\citeauthoryear{Songaila \& Cowie}{Songaila \&
  Cowie}{2010}]{SC2010}
Songaila A.,  Cowie L.~L.,  2010, \mn@doi [ApJ] {10.1088/0004-637X/721/2/1448},
  721, 1448

\bibitem[\protect\citeauthoryear{{Springel}}{{Springel}}{2005}]{Springel2005}
{Springel} V.,  2005, \mn@doi [\mnras] {10.1111/j.1365-2966.2005.09655.x},
  \href {http://adsabs.harvard.edu/abs/2005MNRAS.364.1105S} {364, 1105}

\bibitem[\protect\citeauthoryear{{Stark}, {Ellis}, {Chiu}, {Ouchi}  \&
  {Bunker}}{{Stark} et~al.}{2010}]{Stark2010}
{Stark} D.~P.,  {Ellis} R.~S.,  {Chiu} K.,  {Ouchi} M.,   {Bunker} A.,  2010,
  \mn@doi [\mnras] {10.1111/j.1365-2966.2010.17227.x}, \href
  {http://adsabs.harvard.edu/abs/2010MNRAS.408.1628S} {408, 1628}

\bibitem[\protect\citeauthoryear{{Tilvi} et~al.,}{{Tilvi}
  et~al.}{2014}]{Tilvi2014}
{Tilvi} V.,  et~al., 2014, \mn@doi [\apj] {10.1088/0004-637X/794/1/5}, \href
  {http://adsabs.harvard.edu/abs/2014ApJ...794....5T} {794, 5}

\bibitem[\protect\citeauthoryear{{Vale} \& {Ostriker}}{{Vale} \&
  {Ostriker}}{2004}]{VO2004}
{Vale} A.,  {Ostriker} J.~P.,  2004, \mn@doi [\mnras]
  {10.1111/j.1365-2966.2004.08059.x}, \href
  {http://adsabs.harvard.edu/abs/2004MNRAS.353..189V} {353, 189}

\bibitem[\protect\citeauthoryear{Verner, Ferland, Korista  \& Yakovlev}{Verner
  et~al.}{1996}]{Verner96}
Verner D.~A.,  Ferland G.~J.,  Korista K.~T.,   Yakovlev D.~G.,  1996, ApJ,
  465, 487

\bibitem[\protect\citeauthoryear{{Weinberg}, {Hernsquit}, {Katz}, {Croft}  \&
  {Miralda-Escud{\'e}}}{{Weinberg} et~al.}{1997}]{Weinberg1997}
{Weinberg} D.~H.,  {Hernsquit} L.,  {Katz} N.,  {Croft} R.,
  {Miralda-Escud{\'e}} J.,  1997, in {Petitjean} P.,  {Charlot} S.,  eds,
  Structure and Evolution of the Intergalactic Medium from QSO Absorption Line
  System. p.~133 (\mn@eprint {} {arXiv:astro-ph/9709303})

\bibitem[\protect\citeauthoryear{{Wise}, {Demchenko}, {Halicek}, {Norman},
  {Turk}, {Abel}  \& {Smith}}{{Wise} et~al.}{2014}]{Wise2014}
{Wise} J.~H.,  {Demchenko} V.~G.,  {Halicek} M.~T.,  {Norman} M.~L.,  {Turk}
  M.~J.,  {Abel} T.,   {Smith} B.~D.,  2014, \mn@doi [\mnras]
  {10.1093/mnras/stu979}, \href
  {http://adsabs.harvard.edu/abs/2014MNRAS.442.2560W} {442, 2560}

\bibitem[\protect\citeauthoryear{Worseck, Prochaska, Hennawi  \&
  McQuinn}{Worseck et~al.}{2014a}]{Worseck2014a}
Worseck G.,  Prochaska J.~X.,  Hennawi J.~F.,   McQuinn M.,  2014a, arXiv,
  p.~7405

\bibitem[\protect\citeauthoryear{Worseck et~al.,}{Worseck
  et~al.}{2014b}]{Worseck2014}
Worseck G.,  et~al., 2014b, MNRAS, 445, 1745

\bibitem[\protect\citeauthoryear{{Wyithe} \& {Bolton}}{{Wyithe} \&
  {Bolton}}{2011}]{WB2011}
{Wyithe} J.~S.~B.,  {Bolton} J.~S.,  2011, \mn@doi [\mnras]
  {10.1111/j.1365-2966.2010.18030.x}, \href
  {http://adsabs.harvard.edu/abs/2011MNRAS.412.1926W} {412, 1926}

\bibitem[\protect\citeauthoryear{{Wyithe} \& {Loeb}}{{Wyithe} \&
  {Loeb}}{2006}]{WL2006}
{Wyithe} J.~S.~B.,  {Loeb} A.,  2006, \mn@doi [\apj] {10.1086/502620}, \href
  {http://adsabs.harvard.edu/abs/2006ApJ...646..696W} {646, 696}

\bibitem[\protect\citeauthoryear{{Zel'dovich}}{{Zel'dovich}}{1970}]{Zeldovich1970}
{Zel'dovich} Y.~B.,  1970, \aap, \href
  {http://adsabs.harvard.edu/abs/1970A%26A.....5...84Z} {5, 84}

\makeatother
\end{thebibliography}

\end{document}